\documentclass[aps,prd,superscriptaddress,twocolumn,floatfix,letterpaper]{revtex4-2}

\usepackage{anyfontsize}
\usepackage[hidelinks]{hyperref}
\usepackage{graphicx}
\usepackage{amsmath}
\usepackage{bm}
\usepackage{lipsum}
\usepackage{color}
\usepackage{xcolor}
\usepackage{lineno}
\usepackage{float}
\usepackage{bm} 
\newcommand{\eg}{{\it e.g.}}
\newcommand{\ie}{{\rm i.e.}}

\newcommand{\beq}{\begin{equation}} 
\newcommand{\eeq}{\end{equation}}
\newcommand{\bqa}{\begin{eqnarray}}
\newcommand{\eqa}{\end{eqnarray}}
\usepackage{csquotes}
\MakeOuterQuote{"}

\usepackage{float}

\begin{document}

\title{Semiclassical treatment of bottomonium
suppression and regeneration in $p+{\rm Pb}$ collisions}
\author{Sabin Thapa}
\affiliation{Kent State University, Department of Physics, Kent, Ohio 44242 USA}
\author{Biaogang Wu}
\affiliation{Cyclotron Institute and Department of Physics and Astronomy, Texas A\&M University, College Station, TX 77843-3366, USA}
\author{Ramona Vogt}
\affiliation{Nuclear and Chemical Sciences Division, Lawrence Livermore National Laboratory, Livermore, California 94551, USA}
\affiliation{Physics and Astronomy Department, University of California at Davis, Davis, California 95616, USA}
\author{Ralf Rapp}
\affiliation{Cyclotron Institute and Department of Physics and Astronomy, Texas A\&M University, College Station, TX 77843-3366, USA}

\begin{abstract}
We study bottomonium suppression in $p+ {\rm Pb}$ relative to $p+p$ collisions at center-of-mass energies of $\sqrt{s_{NN}}= 5.02$ and 8.16~TeV. Specifically, we combine cold nuclear matter effects (nuclear modifications of the parton densities, energy loss and momentum broadening) with those from hot nuclear matter (suppression and regeneration) by implementing the formation of a quark-gluon plasma in hydrodynamic simulations.  Bottomonium transport in the quark-gluon plasma is evaluated semiclassically, employing two different reaction rates. The first includes quasi-free inelastic scattering and gluo-dissociation employing a perturbative coupling to the medium. The second is based on in-medium $T$-matrix calculations where the input potential is constrained by lattice quantum chromodynamics to extract the bottomonium masses and dissociation rates. These semiclassical results are compared to previous calculations in an open quantum systems approach and to the experimental data.  Predictions for $\chi_b$ suppression at $\sqrt{s_{NN}} = 8.16$~TeV are also presented.   

\end{abstract}

\keywords{bottomonium production, proton-nucleus collisions, nPDF, energy loss, momentum broadening, quarkonium transport, anisotropic hydrodynamics}
\maketitle

\section{Introduction}
High-energy heavy-ion collisions at the Large Hadron Collider (LHC) and the Relativistic Heavy-Ion Collider (RHIC) provide a unique opportunity to recreate the extreme conditions of the early universe moments after the Big Bang. During this time, the universe existed in a color-deconfined state of nuclear matter known as the quark-gluon plasma (QGP). The study of heavy quarkonia, bound states formed by a heavy quark (charm or bottom) and its antiquark, in these collisions offers valuable insights into the nature of nuclear matter. Specifically, the suppression of quarkonia in nucleus-nucleus ($A+A$) and proton-nucleus ($p+A$) collisions, relative to $p+p$ collisions, serves as a versatile probe of the underlying physics. This suppression arises from a combination of effects, typically categorized into contributions from cold nuclear matter (CNM) and hot nuclear matter (HNM). To accurately investigate and understand the properties of the QGP, it is essential to distinguish the suppression mechanisms induced by the hot medium from those originating in cold nuclear matter interactions. Developing a comprehensive framework that simultaneously accounts for both CNM and QGP effects is key to achieving a deeper understanding of the distinct suppression sources.

Inside the QGP, the attractive potential between quarks and antiquarks is expected to be screened, causing a weakening of the binding of heavy quarkonium states, 
facilitating their melting.  The excited states are more susceptible to suppression in heavy-ion collisions than the ground state due to their weaker binding energies. This phenomenon was initially proposed by Matsui and Satz~\cite{Matsui:1986dk} as evidence of QGP formation.
However, it was subsequently realized that the screening effect is only part of a more complex picture of quarkonium kinetics in QCD matter.
First, inelastic reactions with medium constituents generate dissociation widths that can cause suppression of quarkonium states well below their melting temperatures~\cite{Rapp:2008tf,Mocsy:2013syh,Andronic:2024oxz}.  This, in turn, implies that quarkonium states can also be regenerated by freely moving heavy quarks and antiquarks in the QGP~\cite{Braun-Munzinger:2009dzl}. A robust description of quarkonium production in heavy-ion collision thus requires the deployment of quantitative transport approaches with transport parameters that include the in-medium properties of the various quarkonium states.  

Recent lattice QCD (lQCD) computations confirm the presence of large widths while also indicating that the underlying screening of the potential may be substantially weaker than originally expected~\cite{Larsen:2019bwy,Bala:2021fkm}. Quantitative analyses of the lQCD data using the thermodynamic $T$-matrix approach find that the bound states may survive up to higher QGP temperatures than previously assumed~\cite{ZhanduoTang:2023pla}, with dissociation widths that are much larger than their nominal binding energies~\cite{Tang:2025ypa}.  

Semiclassical transport approaches have been developed over the past $\sim$30 years and can provide a relatively robust description of quarkonium observables in heavy-ion collisions from the SPS and RHIC to the LHC~\cite{Rapp:2008tf,Liu:2015izf,Rapp:2025hyf}. It is now well established that strong QGP suppression of charmonia is followed by regeneration reactions that dominate their inclusive yields in semi/-central Pb+Pb collisions at the LHC, driven by a significant number of tens of $c \overline c$ pairs in the fireball. On the other hand, at RHIC and SPS energies, the regeneration contribution is subleading. The role of regeneration is less obvious for bottomonia because, even in central Pb+Pb collisions at the LHC, there is rarely more than one $b\overline b$ pair in the fireball~\cite{Grandchamp:2005yw,Strickland:2011aa,Krouppa:2015yoa, Du:2017qkv,Wolschin:2020kwt}. However, regeneration from a single $b\overline b$ pair may still be significant, especially since less than 1\% of the $b\overline b$ pairs produced in $p+p$ collisions lead to bottomonia formation. 

In addition, the role of quantum effects is not yet fully understood. They are expected to be important not only in the initial formation phase but also in the medium when the bound state formation times become large. Recent work has addressed this issue by developing quantum transport descriptions based on open quantum systems (OQS)~\cite{Katz:2015qja,Brambilla:2016wgg,Blaizot:2017ypk,Yao:2020xzw,Miura:2022arv}, see, \eg, Refs.~\cite{Rothkopf:2019ipj,Akamatsu:2020ypb} for recent reviews.  
One of the main challenges that remains in these approaches is related to the treatment of regeneration from initial wavepackets that do not overlap with the bottomonium states.

The presence of a hot QCD medium is, however, not the only source of modifications of quarkonium production in $A+A$ collisions relative to the $p+p$ baseline. Studies of the charmonium sector have shown that cold nuclear matter effects can play a significant role, such as the modification of parton distribution functions (PDFs) in the nucleus~\cite{Vogt:2022glr,Eskola:2021nhw,Kovarik:2015cma,AbdulKhalek:2022fyi} as well as  energy loss and momentum broadening from initial parton scatterings prior to quarkonium production~\cite{Arleo:2012rs,Arleo:2013zua,Liou:2014rha,Arleo:2014oha}.  (Nuclear absorption of heavy quarkonium on the incoming nucleons is believed to be negligible at LHC energies~\cite{Lourenco:2008sk}.) Another mechanism that could play a role in $p+A$ collisions is quarkonium interactions with comoving hadrons (or partons)~\cite{Gavin:1988hs,Vogt:1988fj,Gavin:1990gm,Capella:1996va,Ftacnik:1988qv,Armesto:1997sa,Capella:2000zp,Capella:2005cn,Capella:2006mb,Capella:2007jv,Ferreiro:2012rq,Ferreiro:2014bia}, which, however, is a final-state effect that is closely related conceptually to dissociation in the medium~\cite{Gavin:1990gm}. 

To disentangle the roles played by both cold and hot matter effects, it is necessary to systematically analyze heavy quarkonium suppression in both $p+A$ and $A+A$ collisions on the same footing.  In the latter, CNM effects are expected to be subleading, resulting in a 20\%-30\% suppression compared to $p+p$ collisions ($R_{AA} \sim 70-80\%$). This level of suppression is insufficient to explain, for example, the large suppression of bottomonium production observed at the LHC~\cite{CMS:2017ycw,Sirunyan:2018nsz,ALICE:2019pox,Acharya:2020kls,Lee:2021vlb,CMS:2020efs,ATLAS:2022exb,CMS:2023lfu}. 
At lower collision energies, however, CNM effects on bottomonium are expected to become more significant (due to the reduced temperature of the medium) for quantitatively understanding the experimental observations made at RHIC~\cite{Strickland:2023nfm,STAR:2013kwk,PHENIX:2014tbe,STAR:2016pof}. On the other hand, in $p+A$ collisions at LHC energies, the CNM suppression of $\Upsilon$(1S) production is understood to primarily arise from nuclear PDF (nPDF) modifications, energy loss, and momentum broadening.  Thus these effects need to be included in studies of the transverse momentum and rapidity dependencies of the suppression observed in such collisions. 

However, experimental data on the excited $\Upsilon$(2S) and $\Upsilon$(3S) bottomonium states indicate a larger suppression than that of the ground state. This difference in suppression is difficult to understand solely in terms of CNM effects which do not differentiate their modifications based on the identity of the final bottomonium state. Comover models, based on final-state interactions, have been successfully invoked to explain these observations~\cite{Capella:1996va,Armesto:1997sa,Capella:2000zp,Capella:2005cn,Capella:2006mb,Capella:2007jv,Ferreiro:2012rq,Ferreiro:2014bia}. However, there is some ambiguity regarding the identity of the comovers in $p+A$ relative to $A+A$ interactions. (Traditionally, comover models were based on suppression by comoving hadrons. However, recently these models have allowed for the possibility that the comovers could also be deconfined partons.  In this  case, one may consider the comover models to be conceptually equivalent to a phenomenological model of interactions within a deconfined QGP.) A more consistent treatment of final-state quarkonium interactions can be achieved by introducing a short-lived QGP, including a subsequent hadronic phase, in $p + {\rm Pb}$ collisions~\cite{Strickland:2018exs,Noronha:2024dtq}.

Quarkonium suppression in $p+A$ collisions due to a short-lived QGP has been investigated in several previous studies. In Ref.~\cite{Du:2018wsj}, a transport model of $J/\psi$ and $\psi$(2S) production combined EPS09 nPDF modifications with hot QGP effects, including regeneration, using a 2+1D fireball model with a first-order phase transition applied in distinct rapidity intervals. Similarly, Ref.~\cite{Dinh:2019ajl} examined $\Upsilon$ suppression by combining nPDF effects with coherent energy loss within a 2+1D ideal hydrodynamic framework. Reference~\cite{Wen:2022utn} studied \(J/\psi\) and $\psi$(2S) production by incorporating EPS09 nPDFs and hot QGP effects using a complex potential model with screening and in-medium decays, coupled to a 2+1D ideal hydrodynamic background with a first-order phase transition applied in distinct rapidity intervals. The authors of Ref.~\cite{Kim:2022lgu} considered only hot QGP effects on $\Upsilon$($n$S) production in several collision systems: $p+p$, $p+ {\rm Pb}$, $p+ {\rm O}$, and ${\rm O} + {\rm O}$, using a transport theory without regeneration, taking dissociation rates calculated assuming gluo-dissociation and inelastic parton scattering. The 2+1D viscous hydrodynamic background was simulated using the SONIC code~\cite{Romatschke:2015gxa}. Similarly, in Ref.~\cite{Chen:2023toz}, the EPS09 nPDFs were included with QGP effects on $\Upsilon$($n$S) production, employing transport theory coupled to 2+1D ideal hydrodynamics at central rapidity. In our prior work~\cite{Strickland:2024oat}, both CNM effects (EPPS21 nPDFs, energy loss and momentum broadening) and QGP effects were employed to explain $\Upsilon$($n$S) suppression in $p+ {\rm Pb}$ collisions.  The in-medium effect was modeled using an open quantum systems approach with potential nonrelativistic QCD effective theory (OQS+pNRQCD) coupled to a 3+1D anisotropic hydrodynamic background.  All these studies found that there could be important final-state effects on heavy quarkonia arising from propagation through a transient QGP.
 
In this work, we present a comprehensive calculation of cold and hot nuclear matter effects on bottomonium production in minimum-bias $p+ {\rm Pb}$ collisions. We employ a three-stage model that accounts for nPDF effects on initial bottomonium production using the EPPS21 nPDFs~\cite{Eskola:2021nhw}; coherent energy loss and transverse momentum broadening following the formalism of Refs.~\cite{Arleo:2012rs,Arleo:2013zua,Liou:2014rha,Arleo:2014oha}; and the real time evolution of bottomonium states in a QGP produced in $p+{\rm Pb}$ collisions using a semiclassical transport approach~\cite{Du:2017qkv}. Two different scenarios for the dissociation rates are employed: the first utilizes in-medium binding energies from $T$-matrix calculations with the internal-energy potential~\cite{Riek:2010fk} combined with a perturbative coupling to a quasiparticle QGP~\cite{Du:2017qkv} while the second employs nonperturbative rates based on self-consistent $T$-matrix calculations constrained by lattice QCD, as recently obtained in Refs.~\cite{Tang:2025ypa,Wu:2025hlf}. In particular, the effect of bottomonium regeneration in $p+A$ collisions is implemented for the first time.
These results are compared with the OQS+pNRQCD effects calculated previously~\cite{Strickland:2024oat,Brambilla:2023hkw}. All QGP-induced suppression effects are evaluated using trajectories obtained from a 3+1D dissipative hydrodynamic background with a realistic equation of state. The magnitude of each effect as a function of transverse momentum and rapidity is systematically calculated, individually and in combination, allowing comparisons with experimental data collected in different rapidity intervals by the ALICE~\cite{ALICE:2019qie}, ATLAS~\cite{ATLAS:2017prf}, CMS~\cite{CMS:2022wfi}, and LHCb~\cite{LHCb:2018psc} collaborations in a manner consistent with the spectra and rapidity dependence of soft hadron production in $p+{\rm Pb}$ collisions.

This paper is organized as follows: In Sec.~\ref{sec:cnm}, our implementation of initial-state CNM effects on bottomonium production is detailed, with a brief overview of the nPDF effects as well as energy loss and transverse-momentum broadening. In Sec.~\ref{sec:TAMU_transport_prim}, the final-state QGP effects on the bottomonium states are elaborated upon, describing how their primordial suppression in a 3+1D hydrodynamically expanding QGP is computed using a semiclassical kinetic rate equation. We present a semiclassical approach to bottomonium transport in the expanding hot QGP medium, focusing on primordial suppression, discussing the two reaction rates used in the calculations, and show results that include CNM and primordial HNM effects with feeddown contributions, comparing the results to experimental data. In Sec.~\ref{sec:reg}, we investigate the potential effect of regeneration of bottomonium states in the short-lived QGP in $p+$Pb collisions at LHC energies. We present results that combine CNM and primordial HNM effects, including feeddown, with regeneration, again comparing to the experimental data. Finally, Sec.~\ref{sec:conclusions} summarizes the findings and presents the outlook for future investigations.

\section{Cold Nuclear Matter Effects}
\label{sec:cnm}
Before the hot QGP is formed in high energy $p+A$ collisions, quarkonium production is modified by the presence of a cold nucleus. In this work, as in Ref.~\cite{Strickland:2024oat}, modifications of quarkonium production by the nuclear medium resulting from modifications of the parton densities in the nucleus and the interaction of the produced quarkonium states with the medium are taken into account.  The modification of the parton distributions in nuclei relative to those of the free proton, often referred to as shadowing~\cite{Eskola:1991ec, Vogt:2015uba}, can change the total quarkonium yields as well as their kinematic distributions.  Once the $Q \overline Q$ pair that produces final state quarkonium has been created in the initial proton-nucleon collisions, it can interact with the surrounding medium, affecting the propagation of the $Q \overline Q$ within the collision zone, resulting in an overall energy loss, with associated momentum broadening~\cite{Arleo:2012rs}.

Here, $\Upsilon$ production is calculated within the color evaporation model (CEM)~\cite{Gavai:1994in}.  While the improved color evaporation model~\cite{Ma:2016exq} may also be applied, resulting in distinct distributions of the $\Upsilon$ states, these relative changes do not affect the nuclear suppression factor.  In addition, the CNM effects included are effectively independent of the $\Upsilon$ state produced.  The nuclear modifications due to the parton densities, referred to as nPDF effects and discussed in Sec.~\ref{subsec:npdf}, depend on the $b \overline b$ pair mass and not that of the final state $\Upsilon$.  The energy loss effect, as implemented in Ref.~\cite{Arleo:2012rs}, does depend on the $\Upsilon$ mass, but, because the differences between the $\Upsilon$ masses are not large relative to the bottom quark mass, an average $\Upsilon$ mass will be employed, as discussed in Sec.~\ref{subsec:eloss}.

\subsection{Nuclear PDF (nPDF) Modifications}
\label{subsec:npdf}

The $\Upsilon$ production cross section in $p+p$ collisions in the CEM is
\begin{eqnarray}
\sigma_{\rm CEM}(pp) & = & F_C \sum_{i,j} 
\int_{4m^2}^{4m_H^2} d\hat{s}
\int dx_1 \, dx_2~ \label{sigCEM} \\ 
& & \hspace{-8mm} \times  F_i^p(x_1,\mu_F^2,k_{T_1})~ F_j^p(x_2,\mu_F^2,k_{T_2})~ 
\hat\sigma_{ij}(\hat{s},\mu_F^2, \mu_R^2) \, , \nonumber
\end{eqnarray}
where $i$, $j$ represent the incoming partons initiating $b \overline b$ production from processes $i+j= g+g$, $q + \overline q$, and, at next to leading order (NLO), $q(\overline q)+ g$, contributing to the partonic cross section $\hat\sigma_{ij}(\hat {s},\mu_F^2, \mu_R^2)$.  The cross section depends on the quark mass, factorization scale $\mu_F$ and renormalization scale $\mu_R$.   The NLO heavy-flavor cross section is obtained using the HVQMNR code~\cite{Mangano:1991jk}. The values for the bottom quark
mass, $m$, and the scales $\mu_F$ and $\mu_R$,
are determined from a fit to the total $b \overline b$
cross section at NLO:
$(m,\mu_F/m_T, \mu_R/m_T) = (4.65 \pm 0.09 \, {\rm GeV}, 1.40^{+0.77}_{-0.59}, 1.10^{+0.22}_{-0.20})$.
The scales are defined relative to the transverse mass of the $b \overline b$ pair,
\mbox{$\mu_{F,R} \propto m_T = \sqrt{m^2 + p_T^2}$}, where 
the $p_T$ is the transverse momentum of the $b \overline b$ pair, 
\mbox{$p_T^2 = 0.5(p_{T,Q}^2 + p_{T,{\overline Q}}^2)$}.

The normalization factor $F_C$ is obtained by fitting the energy dependence of the summed $\Upsilon$($n$S) cross sections at $y = 0$ measured through their decays to lepton pairs.  A summed branching ratio was employed because early $\Upsilon$ measurements did not have sufficient mass resolution to differentiate among the three $\Upsilon$ peaks in the dilepton spectrum.  The normalization factors of individual bottomonium states are determined based on the individual branching ratios to dileptons and feeddown from excited states.
  
The parton densities in Eq.~\eqref{sigCEM} include intrinsic $k_T$ broadening.
These parton densities are assumed to factorize into the normal collinear parton densities and a $k_T$-dependent function, $F^p(x,\mu_F^2,k_T) = f^p(x,\mu_F^2)G_p(k_T)$.
The CT10 proton parton distribution functions (PDFs)
\cite{Lai:2010vv}, used in the determination of $F_C$, are employed in the calculations of $f^p(x,\mu_F^2)$.

The $k_T$ broadening is introduced to keep the $p_T$ distribution finite as $p_T \rightarrow 0$ because, at leading order (LO), the transverse momentum of the $Q \overline Q$ pair is zero.  Broadening has typically been modeled by adding an intrinsic transverse momentum, $k_T$, to the parton densities.  This broadening plays the role of low transverse momentum QCD resummation \cite{Lo:1979he} without employing it directly.  A $k_T$ broadening was initially added to the quarks in the initial state for Drell-Yan production \cite{Lo:1979he} to create low $p_T$ lepton pairs because the Drell-Yan process also results in zero $p_T$ lepton pairs at leading order.  

In the HVQMNR code, an intrinsic $k_T$ is added to each final state bottom quark, 
rather than to the initial state \cite{Mangano:1991jk}.  The $k_T$ broadening was introduced in Ref.~\cite{Mangano:1992kq} to compensate for the strong effect of fragmentation on fixed-target production of charm quarks. 
In the initial state, the intrinsic $k_T$ function multiplies the parton
distribution functions for both hadrons, 
assuming the $x$ and $k_T$ dependencies factorize.  
At LO, there is no difference between adding a $k_T$ kick in the initial-state (the partons) or on the 
final-state (produced bottom quarks).  
However, at NLO, when there is a
light parton in the final state, the correspondence is approximate.  
The difference between the two implementations is small if
$\langle k_T^2 \rangle \leq 2-3$ GeV$^2$ \cite{Mangano:1992kq}.

A Gaussian distribution is employed for the broadening effect, $ G_p(k_T) = \exp(-k_T^2/\langle k_T^2
\rangle_p) / {\pi \langle k_T^2 \rangle_p} $. 
It is applied by boosting the transverse momentum
of the $b \overline b$ pair
(plus light parton at NLO)
to its rest frame
from the longitudinal center-of-mass frame.
The transverse momenta of the incident partons, $\vec k_{T \, 1}$ and
$\vec k_{T\, 2}$, or, in this case, the final state $b$ and $\overline b$ quarks,
are redistributed isotropically with unit modulus to satisfy momentum conservation.
Once boosted back to the initial frame, the transverse momentum of
the $b \overline b$ pair changes from $\vec p_T$ to
$\vec p_T + \vec k_{T 1} + \vec k_{T 2}$~\cite{Frixione:1994nb}.  
The rapidity distributions are independent of the intrinsic $k_T$.

The broadening effect 
decreases with increasing center of mass energy because the calculated average
$p_T$ of the pair $b \overline b$ also increases
with $\sqrt{s_{NN}}$.  The value of $\langle k_T^2 \rangle_p$ is assumed to
increase with $\sqrt{s_{NN}}$,  $\langle k_T^2 \rangle_p = [ 1 + (1/n) \ln (\sqrt{s_{NN}}/20)]$~GeV$^2$, with $n = 3$ for $\Upsilon$ production.  Thus, at the LHC $p+A$ energies, 
\mbox{$\langle k_T^2 \rangle_p = 2.84$~GeV$^2$} and 3~GeV$^2$ for $\sqrt{s_{NN}} = 5.02$ and 8.16~TeV,
respectively, within the range
of applicability proposed in Ref.~\cite{Mangano:1992kq}.

The $\Upsilon$ production cross section in $p+A$ collisions is modified as 
\begin{eqnarray}
\sigma_{\rm CEM}(pA) & = & F_C \sum_{i,j} 
\int_{4m^2}^{4m_H^2} d\hat{s}
\int dx_1 \, dx_2~ \label{sigCEM_pA} \\ 
& & \hspace{-15mm} \times F_i^p(x_1,\mu_F^2,k_{T \, 1})~ F_j^A(x_2,\mu_F^2,k_{T \, 2})~ 
\hat\sigma_{ij}(\hat{s},\mu_F^2, \mu_R^2) \, .\nonumber
\end{eqnarray}
Here the parton densities in the nucleus, $F_j^A(x_2,\mu_F^2,k_{T \, 2})$, are taken to be the product of the nuclear modification factor $R_j(x_2,\mu_F^2,A)$ and the free parton density $F_j^p(x_2,\mu_F^2,k_T)$ as,
\bqa
F_j^A(x_2,\mu_F^2,k_{T \, 2}) & = & R_j(x_2,\mu_F^2,A) F_j^p(x_2,\mu_F^2,k_{T \, 2}) \, .
\eqa
No additional $k_T$ broadening is included due to the presence of the nucleus.  Instead, transverse momentum broadening is included, along with energy loss, in the following section.

The NLO EPPS21~\cite{Eskola:2021nhw} nPDF set is used here to calculate the nPDF effect.  The uncertainty on this result is obtained by calculating the cross sections using the central set and the 48 error sets (based on 24 fit parameters), added in  quadrature, as also shown in Ref.~\cite{Strickland:2024oat}.  Note that in other calculations in this approach, such as Ref.~\cite{Vogt:2022glr}, momentum broadening in the nucleus is introduced in the Gaussian distribution, replacing $G_p(k_T)$ with $G_A(k_T)$.  

\subsection{Energy loss and momentum broadening}
\label{subsec:eloss}

Once the proto-quarkonium states are produced in the initial proton-nucleon collisions, they can undergo elastic scattering via gluon exchange with the medium, inducing coherent radiation. These effects are distinct from the gluon radiation resummed in leading-twist factorization and fragmentation functions and must thus be considered explicitly~\cite{Arleo:2012rs,Liou:2014rha}. In $p+A$ collisions, such energy loss and momentum broadening can be quantified by formulating the quarkonium double differential cross section in terms of a rapidity shift, $\delta y$, and transverse momentum broadening, $\delta p_T$, as described in Refs.~\cite{Arleo:2013zua, Strickland:2024oat},
\bqa
\frac{1}{A}\frac{d\sigma_{pA}^{\Upsilon}}{d y \, d^2 p_T}\left(y, p_T \right)  &=& \int_\varphi \int_0^{\delta y_{\rm max}(y)} d \delta y \,\hat{\cal P}(e^{\delta y} - 1, \ell^2_A) \nonumber \\
&& \hspace{-2mm} 
\times \frac{d\sigma_{pp}^{\Upsilon}}{dy \, d^2p_T} \left( y + \delta y, |p_T - \delta p_T | \right) \, ,
\label{eq:elossfinal}
\eqa
where $y$ and $p_T$ are the bottomonium rapidity and transverse momentum respectively; the integral over $\varphi$ is defined as \mbox{$\int_\varphi = \int_{0}^{2\pi} d\varphi/(2\pi)$}; and \mbox{$m_T = \sqrt{p_T^2 + M_\Upsilon^2}$} is the transverse mass of the bottomonium state where $M_\Upsilon = 9.95$~GeV is the average mass of $\Upsilon$(1S), $\Upsilon$(2S), and $\Upsilon$(3S) states \cite{Strickland:2024oat}.  The maximum rapidity shift is $\delta y_{\rm max} (y) = \text{min}(\ln 2, y_{\rm max}-y)$ where \mbox{$y_{\rm max} = \ln(\sqrt{s_{NN}}/m_T)$} is the maximum bottomonium rapidity in the $p+A$ center-of-momentum frame. The magnitude of the transverse momentum broadening is $\delta p_T = \sqrt{\ell^2_A - \ell^2_p}$ where $\ell_A$ and $\ell_p$ are the transverse momentum broadening due to the passage of the bottomonium state through a nucleus or a proton, respectively. The transverse momentum broadening acquired while the pair traverses the target nucleus $A$, as a function of the transport coefficient $\hat{q}$ and the effective path length $L_A$ through the target, is $\ell_A = \sqrt{\hat{q} L_A} $.   Similarly, $\ell_p = \sqrt{\hat{q} L_p}$ with $L_{\rm Pb} = 10.41$~fm and $L_p = 1.5$~fm \cite{Strickland:2024quantum}. The transport coefficient $\hat{q}$ is determined from fits to HERA data~\cite{Golec-Biernat:1998zce},
\begin{eqnarray}
\hat{q} & = & \hat{q}_0 (10^{-2}/x_A)^{0.3} \, , \nonumber \\
x_A & = & {\rm min}(x_0,x_2) \, , \nonumber \\
x_0 & = & (2 m_p L_A)^{-1} \, , \nonumber \\
x_2 & = & (m_T/\sqrt{s_{NN}}) \exp(-y)  \, .
\label{eq:eqs}
\end{eqnarray}
The value of the parameter \mbox{$\hat{q}_0 = \hat{q}(x=10^{-2})$}, $\hat{q}_0 = 0.075 \; {\rm GeV}^2/{\rm fm} $, was determined in Ref.~\cite{Arleo:2012rs}.

In Eq.~(\ref{eq:elossfinal}), $\hat{\cal P}$ is the quenching weight that determines the probability for the energy loss and momentum broadening. It can be written in terms of the dilogarithm function ${\rm Li}_2$ as~\cite{Arleo:2013zua} 
\bqa
\hat{\cal P}(\delta x,\ell_A^2) &=&  \frac{\partial}{\partial \delta x} \, \exp \left\{\frac{N_c \alpha_s}{2 \pi} \left[ {\rm Li}_2\!\left(\frac{-\ell_A^2}{(\delta x)^2 m_T^2} \right) \right. \right.
\nonumber \\
&& \hspace{1.2cm} \left. \left.\hspace{1.5cm} - {\rm Li}_2\!\left(\frac{-\Lambda_{p}^2}{(\delta x)^2 m_T^2} \right) \right] \right\} \, , \;\;\;\;\;
\label{eq:quenching-dilog}
\eqa
where $N_c = 3$ is the number of colors 
and $\delta x$ is the fractional loss, $\delta x = e^{\delta y}-1$ in Eq.~(\ref{eq:elossfinal}).  
The numerator of the second term in Eq.~(\ref{eq:quenching-dilog}), \mbox{$\Lambda_{p}^2 = {\rm max}(\Lambda_{\mathrm{QCD}}^2,\ell_{p}^2)$}, takes the value of either the square of momentum broadening in the proton, $\ell_p^2$, or the square of the QCD scale parameter, $\Lambda_{\rm QCD}^2$.
\begin{figure*}[!hbt]
\includegraphics[width=\textwidth,height=4cm,keepaspectratio]{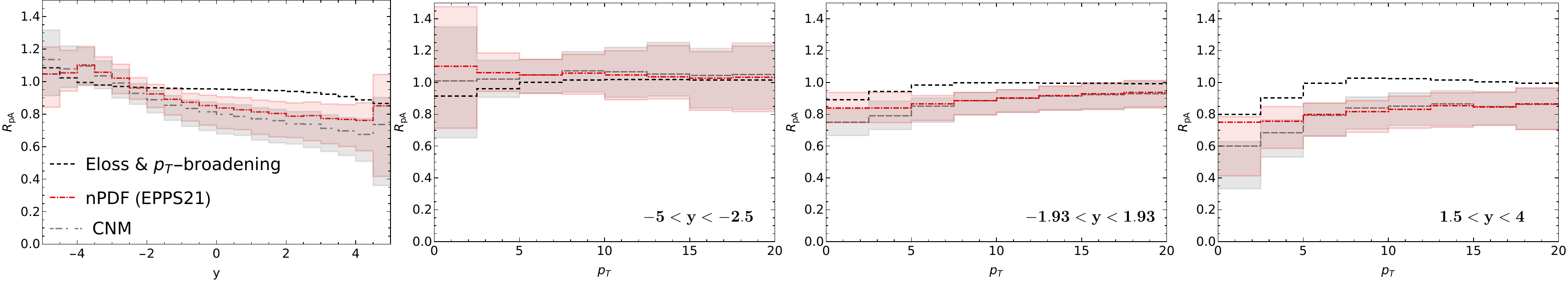}
\caption{The nuclear suppression factor in $\sqrt{s_{NN}} = 8.16$~TeV $p+{\rm Pb}$ collisions due to CNM effects. The nPDF effects are shown in the pink band, including the EPPS21 uncertainties, while the effect of energy loss and transverse momentum broadening is given by the black dashed curve.  The total effect, including both nPDF effects and energy loss plus broadening, is given by the gray band.  The results for $b \overline b$ production are shown as a function of rapidity on the far left while the following three panels show the results as a function of $p_T$ at backward rapidity (center left), midrapidity (center right) and forward rapidity (far right).
} 
\label{fig:rpA-cnmonly-8tev}
\end{figure*}

As in Ref.~\cite{Strickland:2024oat}, we employ the four-loop value of $\alpha_s$ in Eq.~(\ref{eq:quenching-dilog}), evaluated at $\Lambda_{\rm QCD} =\delta p_T$ with $\Lambda_{\rm QCD} = 0.308$~GeV to give $\alpha_s(1.5\;{\rm GeV}) = 0.326$, the value extracted from lattice QCD calculations of the static energy~\cite{Bazavov:2012ka}.  The original formulation of the quenching weights \cite{Arleo:2012rs,Arleo:2014oha} assumed the fixed coupling $\alpha_s = 0.5$.  We have determined that the difference between these choices is small and has no effect on the results of the calculation since $R_{pA}$ due to energy loss begins to differ in the two assumptions at $|y|>3$ and never more than 10\% in the measured range, considerably smaller than the uncertainty band for the nPDF contribution.

We adopt the parametrization of the double differential prompt $\Upsilon$ cross section in Eq.~\eqref{eq:elossfinal} \cite{Arleo:2013zua},
\beq
\frac{d^3 \sigma^{\Upsilon}_{pp}}{d y \, d^2 {p_T}} = \mathcal{N} \left( \frac{p_0^2}{p_0^2 + p_T^2} \right)^\upsilon \left( 1- \frac{2m_T}{\sqrt{s_{NN}}} \cosh y \right)^\beta \, .
\label{eq:ppdist}
\eeq
The normalization $\mathcal{N}$ is irrelevant for the nuclear suppression factors calculated here. The other parameters in Eq.~(\ref{eq:ppdist}) are \mbox{$p_0 = 6.6$ GeV}, $\upsilon = 2.8$, and $\beta = 13.8$, obtained from a global fit of $\Upsilon$ production data~\cite{Arleo:2013zua}. This analytic form, in agreement with the shape of the CEM cross section, is more efficient in numerical calculations as well as useful when sampling bottomonium kinematics when studying their propagation through the hot QGP medium, as detailed in the following sections.

\subsection{Combined cold matter suppression}
\label{subsec:CNMtotal}

Here the combined effects of nPDF modification with energy loss and momentum broadening are presented.
The results for CNM effects alone at $\sqrt{s_{NN}} = 8.16$~TeV are shown in Fig.~\ref{fig:rpA-cnmonly-8tev}. The EPPS21 effect alone is given by the pink bands which include the positive and negative uncertainties on the 24 parameter pairs, added in quadrature.  The combined energy loss and $p_T$ broadening result is represented by the dashed curves because no uncertainties are given on the parameters in the energy loss model.  The full CNM result, the product of the two effects, is given by the gray band.

The results are presented as the nuclear modification factor, $R_{pA}$, defined here as the cross section per nucleon in $p+{\rm Pb}$ collisions relative to the $p+p$ cross section. The nuclear modification factor for the bottomonia states is defined as the ratio of the yields in $p+A$ collisions relative to those in $p+p$ collisions, scaled by the number of nucleon-nucleon collisions $N_{\text{coll}}$, and can be written as,  
\beq
R^{\Upsilon}_{pA} (y, p_T) =  \frac{1}{N_{\rm coll}}\frac{d^3 N^{\Upsilon }_{pA}/d y \, d^2 {\vec p}_T}{d^3 N^{\Upsilon}_{pp}/d y \, d^2 {\vec p}_T} \, , 
\eeq
where $d^3 N^{\Upsilon}_{pp}/d y \, d^2 {\vec p}_T$ is the double differential $\Upsilon$ distribution in $p+p$ given in Eq.~\eqref{eq:ppdist}. While here we have indicated the dependence of nuclear modification factor as a function of $p_T$ and rapidity, we note that it can also be characterized by system quantities such as collision centrality or multiplicity.  In this work, we focus on its kinematic dependence. 

One can define separate modification factors for the nPDF effects and energy loss with momentum broadening because the two components factorize in this calculation,
\beq
R^{\rm CNM}_{pA} = R^{\rm nPDF}_{pA} 
 \times R^{\rm eloss, broad}_{pA} \, .
\label{eq:rpa_CNM}
\eeq
The factor $R^{\rm nPDF}_{pA}$ accounts for the suppression due to nPDF effects while $R^{\rm eloss, broad}_{pA}$ represents the suppression due to coherent energy loss and momentum broadening.  

The results shown in Fig.~\ref{fig:rpA-cnmonly-8tev} do not include any mass or scale variations in Eqs.~(\ref{sigCEM}) and (\ref{sigCEM_pA}) because we assume the mass and scale variations  between the $p+p$ and $p+A$ calculations to be correlated \cite{Vogt:2015uba}.  Of course, taken separately, the mass and scale uncertainties are larger than the nPDF effects in quadrature.

We briefly comment on the combination of nPDFs with energy loss effects.  The nPDFs are extracted from a variety of QCD processes, both in lepton-nucleus and in proton-nucleus collisions.  The gluon distribution in the nucleus is typically only constrained by sum rules and QCD evolution.  However, studies have shown that the production of open-charm mesons and charmonia in $p+A$ collisions at LHC energies provides a significant constraint on the small $x$ gluon distribution, see {\it e.g.}\ \cite{Eskola:2021nhw}. Because these processes are also sensitive to energy loss  effects \cite{Arleo:2012rs,Arleo:2013zua,Arleo:2025oos},  these effects are thus absorbed into the nPDFs.  However, not including the charm measurements would weaken the gluon nPDF constraints.  

A recent paper \cite{ArleoAvez:2023} incorporated LO $J/\psi$ production in $p+A$ collisions in the CEM in calculations with several nPDF sets. They found the gluon nPDF ratio, $R_g$, using the $J/\psi$ data with energy loss included is higher than when energy loss is not included.  See also Ref.~\cite{Arleo:2025oos}.  Thus not including energy loss effects may lead to a lower central value of the gluon nPDF, most important for quarkonium production at the LHC.  There is some kinematics mismatch in the LO approach since the nPDFs are extracted at next-to-leading order and the $J/\psi$ transverse momentum cannot be incorporated at LO.

We note also that the gluon ratio for EPPS21, which included LHC $D$ meson data in the analysis \cite{Eskola:2021nhw}, is quite similar to that of EPPS16 \cite{Eskola:2016} which did not.  It is also worth mentioning that the value of $\hat{q}_0$ in Ref.~\cite{Arleo:2012rs}, used here, did not incorporate nPDF.  However, when saturation effects were included with the same $\hat{q}_0$, the value of $R_{pA}$ is decreased for $J/\psi$ production while the effect on $\Upsilon$ production was negligible. 

The rapidity dependence is given in the left-most panel of Fig.~\ref{fig:rpA-cnmonly-8tev}.  At the most negative rapidities the nPDF effects show evidence of antishadowing at the relatively large $x_2$ accessed in this region while low $x_2$ shadowing is evident at forward rapidity.  The results are presented at $\sqrt{s_{NN}} = 8.16$~TeV; the antishadowing peak would be slightly more evident for $\sqrt{s_{NN}} = 5.02$~TeV since $x_2$ is $\sim 60$\% larger at the lower energy.  Note that the uncertainty band is larger for the most forward and backward rapidity bins due to a degradation of statistics as the $\Upsilon$ cross section falls off as a function of rapidity.  The effects of energy loss and momentum broadening are not a strong function of rapidity but generally give a suppression factor of slightly less than unity over the entire rapidity range except at $y < -4$.  Thus combining the two effects leads to slightly stronger cold matter suppression than that due to nPDF effects alone.

The other three panels of Fig.~\ref{fig:rpA-cnmonly-8tev} show $R_{pA}$ as a function of $p_T$ at negative rapidity, midrapidity, and forward rapidity respectively.  As the rapidity interval moves from backward to forward, corresponding to a decrease in momentum fraction $x_2$, shadowing becomes more apparent, and is typically largest at low $p_T$.  In the backward rapidity region, the ratio $R_{pA}$ fluctuates around unity with the uncertainty band generally becoming wider as $p_T$ increases, aside from the lowest $p_T$ bin.  The HVQMNR code, a negative weight Monte Carlo, can suffer from incomplete numerical cancellation of divergences at low $p_T$.  Any influence of the negative weights can generally be overcome by the introduction of the Gaussian intrinsic $k_T$ kick, as described in Sec.~\ref{subsec:npdf}.  This issue is common to all rapidity bins but the effect is enhanced in the backward region where antishadowing is more prominent and small opposing fluctuations may be enhanced.  At midrapidity and forward rapidity, in the range where the shadowing ratio is less than unity over the entire rapidity range covered, all the nPDF sets produce an overall shadowing effect that is strongest at low $p_T$ and more forward rapidity.

The combination of energy loss and $p_T$ broadening appears to be a weaker function of $p_T$ than the nPDF modification alone.  It is most visibly different from unity at low $p_T$ and forward rapidity, becoming consistent with $R_{pA}^{\rm eloss,broad} \sim 1$ at $p_T \sim 7$~GeV.

\section{Quarkonium Suppression in the QGP}
\label{sec:TAMU_transport_prim}
To understand how heavy quarkonium evolves in the hot QGP medium, we consider a semiclassical approach where the kinetic rate equation can be evaluated numerically to obtain the time evolution of bottomonium in the QGP~\cite{Du:2017qkv}. We also compare this result with the quantum mechanical solution obtained by using QTraj-NLO which solves a Lindblad equation for the bottomonia states in the QGP medium obtained within the NLO pNRQCD effective field theory~\cite{Omar:2021kra, Brambilla:2022ynh, Strickland:2024oat}. 

The dynamics of the bottomonium yields in the QGP are given by the semiclassical kinetic rate equation, including loss and gain terms, as in Ref.~\cite{Du:2017qkv}, 
\beq
\frac{d N_{\Upsilon} (\tau)}{d \tau} = - \Gamma_{\Upsilon} (T(\tau)) [N_{\Upsilon} (\tau) - N_{\Upsilon}^{\text{eq}} (\tau)] \, .
\label{eq:rate-eq}
\eeq
The two transport coefficients governing the rates are $\Gamma_{\Upsilon} (T(\tau))$, the temperature and time-dependent inelastic reaction rate, and $N_{\Upsilon}^{\text{eq}} (\tau)$, the equilibrium limit on the number of $\Upsilon$ states. Here, $\Upsilon$ represents a generic bottomonium state.
The rate equation can be decomposed into two parts: primordial suppression (first term in the right hand side of Eq. \eqref{eq:rate-eq}) and regeneration (second term). In this section we will focus on primordial suppression, leaving the discussion of the regeneration to Sec.~\ref{sec:reg}.

The number of primordial $\Upsilon$ states produced in the initial $p+A$ collisions is 
\beq
N_{\Upsilon}^{\text{prim}} (\tau_i) = N_{\text{coll}} \frac{\sigma^{\text{tot}}_{pN \xrightarrow{} \Upsilon}}{\sigma^{\text{in}}_{pN}} \, , \label{eq:N_ups_pp} 
\eeq
where the total production cross section for a given $\Upsilon$ state can be calculated as in Eq.~(\ref{sigCEM}).  Here $\sigma^{\rm in }_{pN}$ is the inelastic cross section at the appropriate center-of-mass energy, 67.6~mb at $\sqrt{s} = 5.02$~TeV and 71~mb at 8.16~TeV.  The number of proton-nucleon collisions, $N_{\rm coll}$, can be calculated using the optical Glauber model~\cite{Wiedemann:1143387}.  The number of primordially-produced $\Upsilon$ states will be less than unity at LHC energies.

Primordial $\Upsilon$ suppression in the QGP is given by
the first term in Eq.~\eqref{eq:rate-eq}
\beq
\frac{d N_{\Upsilon}^{\text{prim}} (\tau)}{d \tau} = - \Gamma_{\Upsilon} (T(\tau)) N_{\Upsilon}^{\text{prim}} (\tau) \, .
\label{eq:rate-eq-prim}
\eeq
Integrating Eq.~\eqref{eq:rate-eq-prim} from the initial formation time, $\tau_i$, of the $\Upsilon$ state to the freeze-out time $\tau_f$, at the decoupling temperature $T_{\rm dec}$, gives the number of primordially-produced bottomonia surviving the QGP,
\beq
N_{\Upsilon}^{\text{prim}}(\tau_f) = N_{\Upsilon}^{\text{prim}} (\tau_i) \exp(-\int_{\tau_i}^{\tau_\text{f}}\Gamma_{\Upsilon} (T(\tau')) d \tau')  \, .
\label{eq:rate-eq-prim-sol}
\eeq
%


In this section we outline the ingredients of primordial suppression due to the hot nuclear matter. First, Sec.~\ref{subsec:rxnrates} introduces the reaction rates that govern the suppression of primordially produced bottomonia. Next, Sec.~\ref{subsec:ahydro} summarizes the 3+1D hydrodynamic (aHydroQP) background used in the HNM evolution. In Sec.~\ref{subsec:rpa_hnm} we implement these rates in the medium and present the resulting hot–medium suppression quantified by $R_{pA}^{\mathrm{HNM}}(y,p_T)$. Sec.~\ref{subsec:feeddown} details the treatment of feeddown from excited states and how it is incorporated into the suppression. Finally, Sec.~\ref{subsec:results_wout_reg} shows results and predictions that combine primordial (HNM) suppression with CNM effects for $\Upsilon(nS)$ and $\chi_b(nP)$, explicitly excluding regeneration.

\subsection{Bottomonium reaction rates in a strongly-coupled QGP}
\label{subsec:rxnrates}

The reaction rates in this study are drawn from two TAMU scenarios, both evaluated within the thermodynamic $T$-matrix approach. 

\begin{figure}[h!]
   \centering
    \includegraphics[width=\columnwidth]{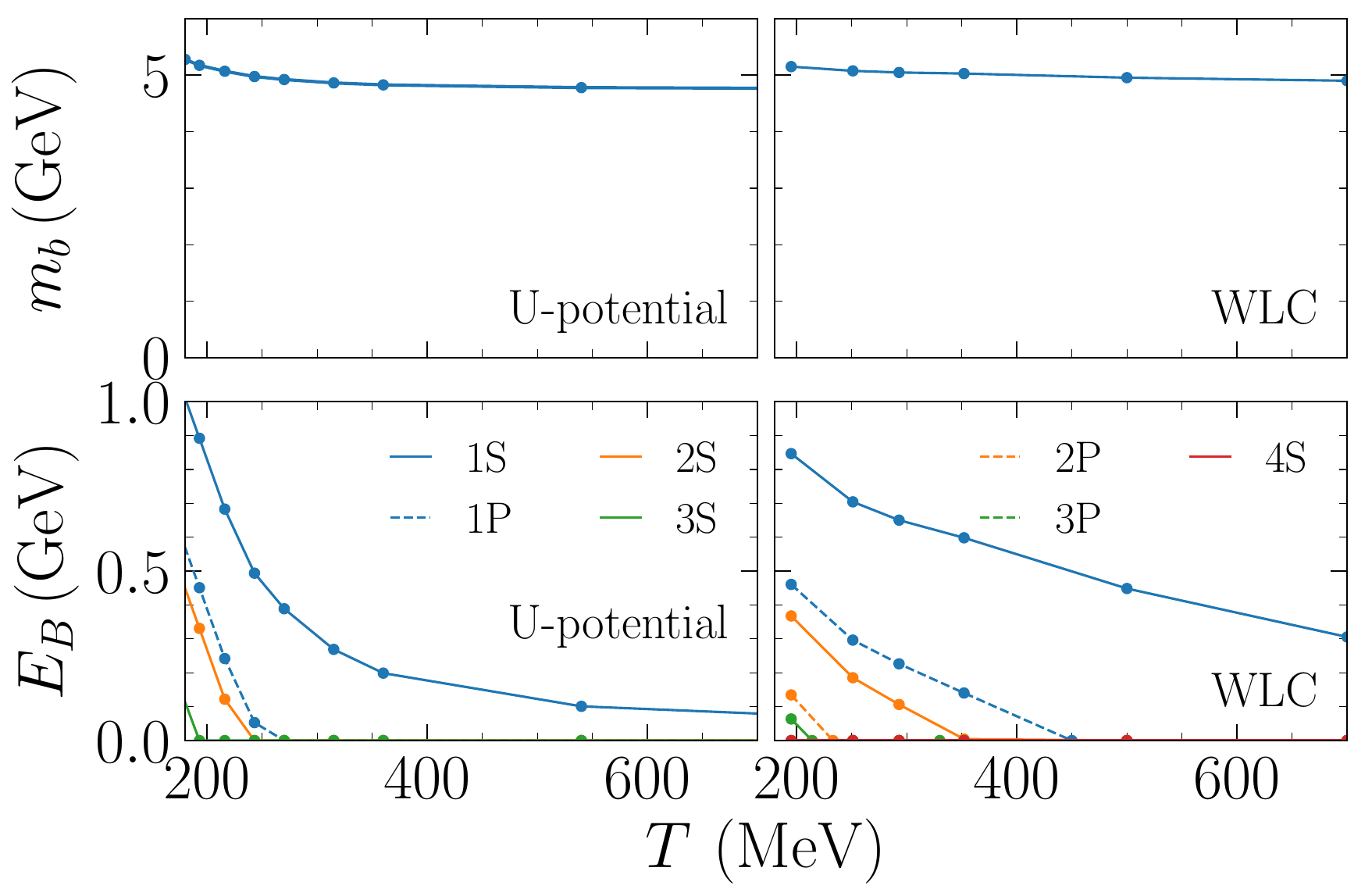}
 \caption{Bottom-quark mass (upper panels) and bottomonium binding energies (lower panels) as a function of temperature for the $U$-potential scenario (left) and WLCs (right) constrained by lattice QCD data.
  In the right panel, solid lines represent the binding energies of S-wave states, the 1S (blue), 2S (orange), 3S (green), and 4S (red) states, while the dashed lines represent the binding energies of P-wave states, the 1P (blue), 2P (orange), and 3P (red) states.
  Only the 1S, 2S, 1P, and 3S states are included in the $U$-potential scenario.
  }
    \label{fig:mass}
\end{figure}

\begin{figure}[h!]
   \centering
   \includegraphics[width=\columnwidth]{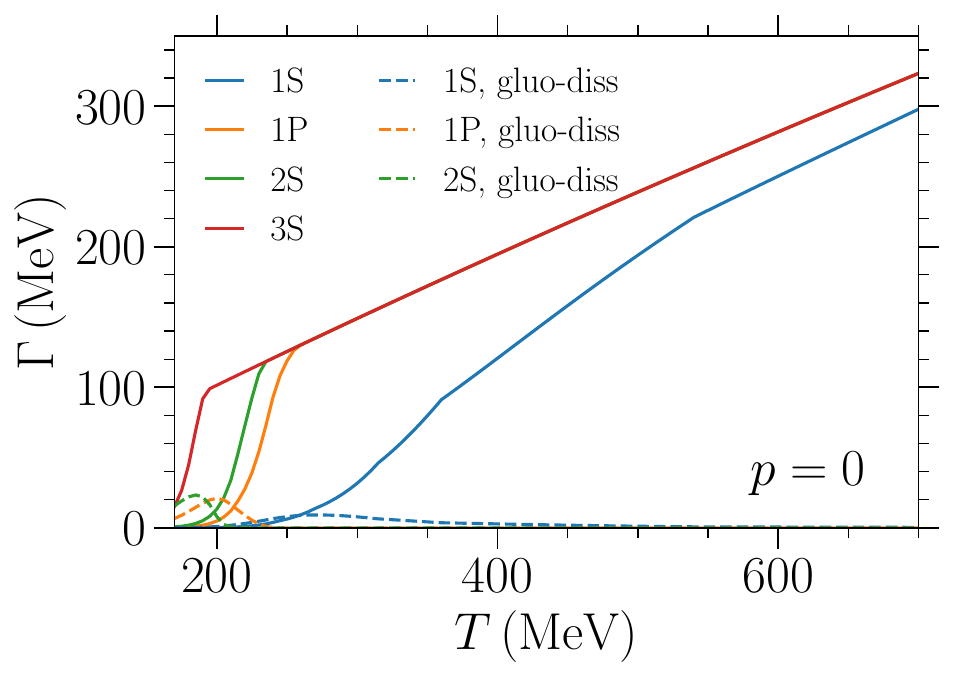}
  \caption{Perturbative rates based on a in-medium $U$-potential for bottomonium binding, as functions of temperature at $p=0$~\cite{Du:2017qkv}. The solid lines represent the inelastic scattering rates while the dashed lines represent the gluo-dissociation rates for the 1S (blue), 1P (orange), 2S (green), and 3S (red) states.
    }
    \label{fig:rates_pert}
\end{figure}

\begin{figure}[ht!]
    \centering
    \includegraphics[width=\columnwidth]{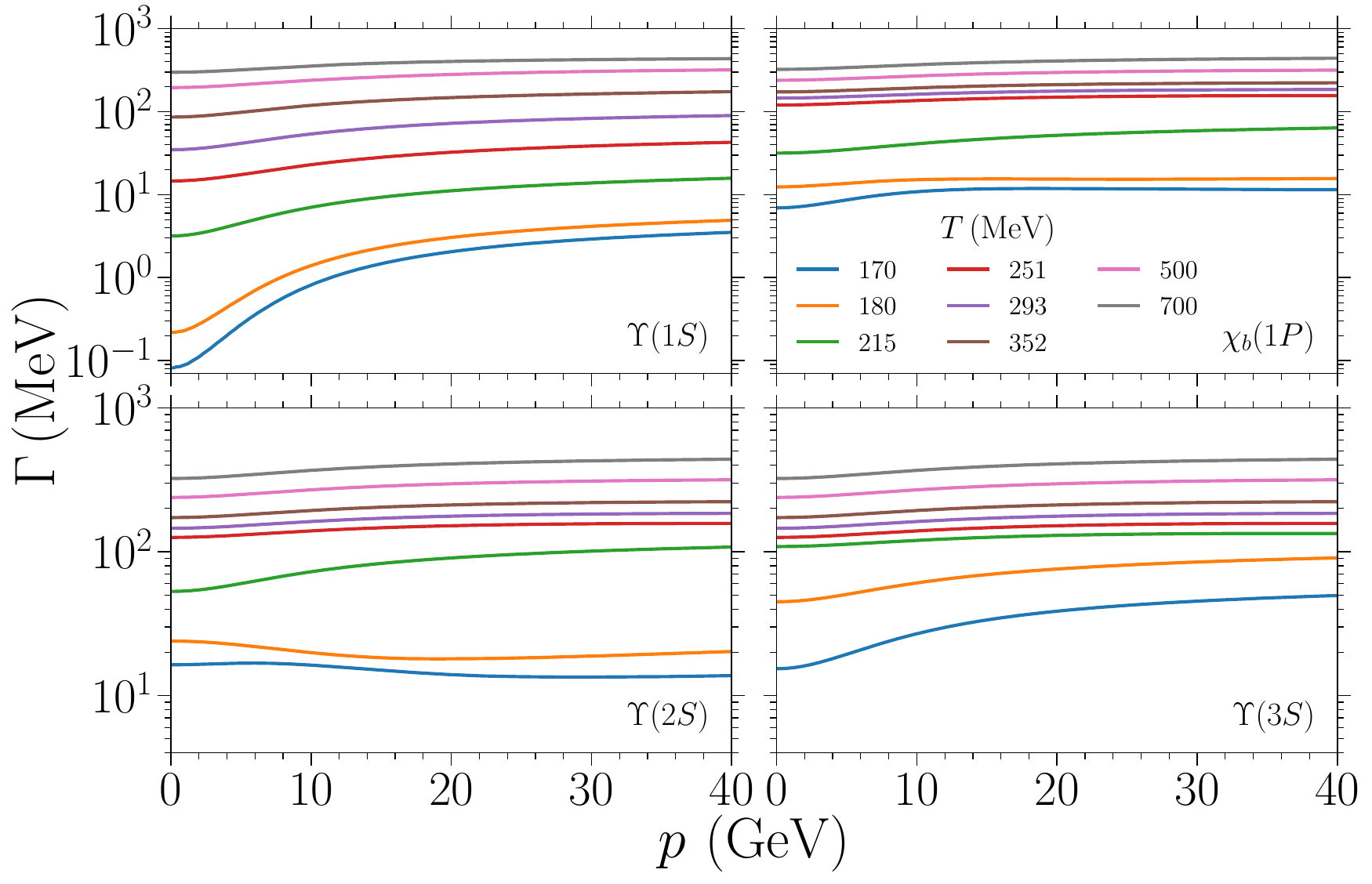}
    \caption{Perturbative bottomonium rates (inelastic scattering and gluon dissociation) assuming an in-medium $U$-potential,
    as functions of momentum for various temperatures~\cite{Du:2017qkv}.
    Each panel corresponds to a specific bottomonium state. The colors represent different temperatures, ranging from $T=170$~MeV (blue) to $T=700$~MeV (gray), as indicated in the legend.
}
    \label{fig:rates_p_pert}
\end{figure}

The first approach, referred to as TAMU-P in the discussion, assumes the finite-temperature internal energy ($U_{Q\bar{Q}}$) as the potential for the quarkonium interaction.  The binding energies and heavy quark masses employed in the interaction were calculated in Ref.~\cite{Riek:2010fk} utilizing constraints from Euclidean quarkonium correlator ratios determined from finite temperature lattice QCD.  The temperature dependence of the bottom quark mass and the bottomonium binding energies are shown on the left-hand side of  Fig.~\ref{fig:mass}.  These quantities were utilized to calculate the quasi-free bottomonium dissociation rates, assuming a perturbative coupling to the thermal medium composed of massive light quark and gluon quasiparticles, for inelastic $i + \Upsilon \rightarrow i + b + \overline b$ scattering ($i = q$, $\overline q$, $g$) and gluo-dissociation, $g+ \Upsilon \to b + \overline b$~\cite{Du:2017qkv}. These rates have been widely used to calculate charmonium~\cite{Du:2017qkv}, bottomonium~\cite{Zhao:2010nk}, and $B_c$~\cite{Wu:2023djn} transport in heavy-ion collisions.  The calculations provide fair agreement with the available data from heavy-ion collisions at the SPS, RHIC and the LHC.
The temperature dependence at $p=0$ is shown in Fig.~\ref{fig:rates_pert} while the momentum dependence of these rates at selected temperatures 
is shown in Fig.~\ref{fig:rates_p_pert}.

The second scenario, referred to as TAMU-NP in the remainder of the text, is a very recent thermodynamic $T$-matrix calculation~\cite{Wu:2025hlf} where the in-medium potential itself is constrained by fits to nonperturbative lattice QCD data, specifically Wilson line correlators (WLCs)~\cite{ZhanduoTang:2023tsd}.
The resulting potential is only weakly
screened, even at rather high temperatures, leading to substantial binding of the bottomonium states well into the QGP.  Using the same input potential, these calculations are embedded into a quantum many-body approach to the equation of state with self-consistently calculated one- and two-parton correlation functions. The off-shell effects from large collisional widths are accounted for through broad parton spectral functions, in connection with interference effects, in a strongly-coupled QGP.
The bare light parton masses are constrained to reproduce lattice QCD data on the equation of state and thus provide a more realistic evaluation
of heavy quark and quarkonium properties. 

The resulting bottomonium spectral functions, while still allowing for bound states, become very broad because the heavy quark widths dominate as the binding is reduced, making it very difficult to extract the bottomonium widths and thus their decay rates. In addition, the 3-momentum dependence of the $Q\overline Q$ 
spectral functions is currently not fully known. Therefore, a novel analysis has been performed by determining the $T$-matrix poles in the complex energy plane~\cite{Tang:2025ypa} to accurately extract the $\Upsilon$ masses and widths.
Since the in-medium bottom quark mass is known, as shown in Fig.~\ref{fig:mass}, one can readily extract the bottomonia binding energies from their pole positions in the complex plane~\cite{Tang:2025ypa}.

The dissociation rates are calculated in a quasi-free approximation to express the rates in terms of nonperturbative heavy-light scattering amplitudes, $\mathcal{M}$,  and the heavy quark and light parton spectral functions, $\rho_{Q}$ and $\rho_{p}$, respectively.  The momentum dependence, $\Gamma(\vec p)$, is obtained from the same $T$-matrix approach~\cite{Wu:2025hlf},
%
\begin{eqnarray}
   \Gamma\left(\vec{p}\right) & = & \sum_i \frac{2}{2 \varepsilon_Q(\vec{p})} \int \frac{d \omega^{\prime} d^3 \vec{p}^{\prime}}{(2 \pi)^3 2 \varepsilon_Q\left(\vec{p}^{\prime}\right)} \frac{d \nu d^3 \vec{q}}{(2 \pi)^3 2 \varepsilon_i(\vec{q})} \nonumber \\
    & \times &\frac{d \nu^{\prime} d^3 \vec{q}^{\prime}}{(2 \pi)^3 2 \varepsilon_i\left(\vec{q}^{\prime}\right)}
   \delta^{(4)}\left( p+q-p'-q' \right) \nonumber \\
   & \times &\frac{(2 \pi)^4}{d_Q} \sum_{a, l, s} |\mathcal{M}|^2 \rho_Q\left(\omega^{\prime}, \vec{p}^{\prime}\right) \rho_i(\nu, \vec{q}) \rho_i\left(\nu^{\prime}, \vec{q}^{\prime}\right) \nonumber \\
   & \times &\left[1 - n_Q\left(\omega^{\prime}\right)\right] \left[1 - n_{\bar{Q}}\left(\varepsilon_{\bar{Q}}\right)\right] n_i(\nu) \left[1 \pm n_i\left(\nu^{\prime}\right)\right] \nonumber \\ 
   & \times &\left[ 1 - e^{i \vec{k} \cdot \vec{r}} \right] \label{eq:rate} \, .
\end{eqnarray}
%
Here, $d_Q$ is the heavy-quark degeneracy factor; $\vec{p}$, $\vec{p}^{\prime}$, $\vec{q}$, and $\vec{q}^{\prime}$ are the 
3-momenta of the incoming heavy quark, outgoing heavy quark, incoming light parton, and outgoing light parton, respectively;
$\omega^{\prime}$, $\nu^{\prime}$ and $\nu$ are the outgoing heavy quark, light parton and incoming light parton energies; and
$\varepsilon_Q$ and $\varepsilon_p$ are the corresponding on-shell energies.
The quantities $n_{Q}$ and $n_p$ denote the distribution functions of heavy quarks and light partons while 
$\vec k = \vec{p} - \vec{p}^{\prime}$ is the momentum transfer.
Relative to earlier calculations employing the perturbative coupling, the quasi-free approximation was significantly amended by including quarkonium 
wavefunction effects through an interference factor, $1 - \exp(i \vec{k} \cdot \vec{r})$, the last term in Eq.~\eqref{eq:rate}. This factor was previously derived in a perturbative approximation~\cite{Laine:2006ns} and has the effect of generating an imaginary part of the heavy quark potential. It suppresses 
the dissociation rates most strongly for compact bound states, \ie, for smaller radii $r$. This effect can be interpreted as the 
interference of the scattering
amplitudes off the heavy quark and antiquark, leading to a rate vanishing for $r\to 0$. 
%

\begin{figure}[h!]
   \centering
   \includegraphics[width=\columnwidth]{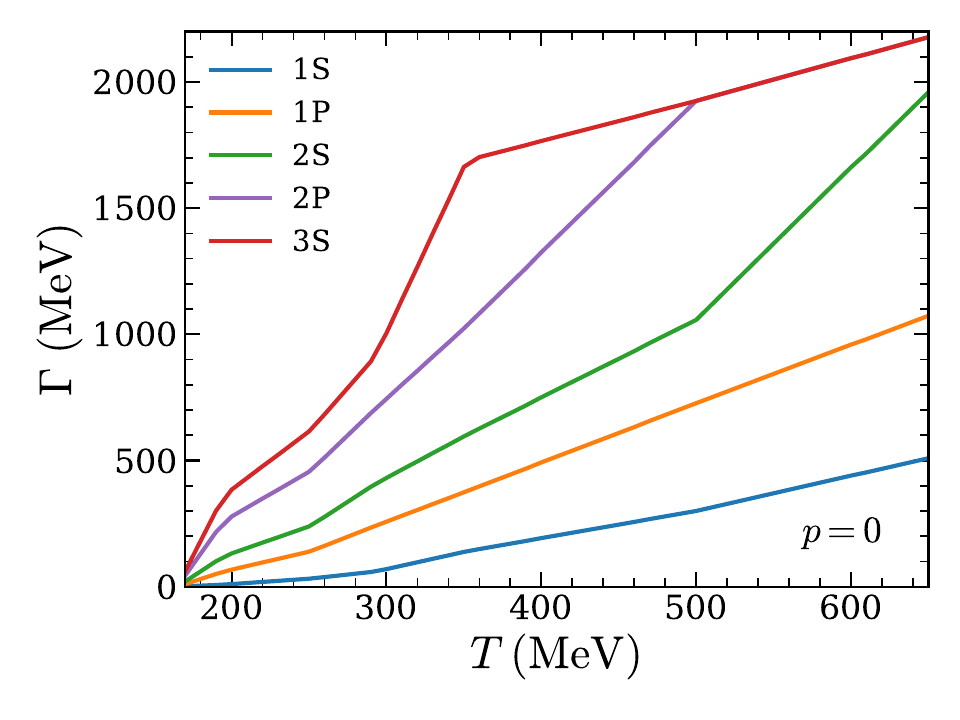}
   \caption{Nonperturbative bottomonium dissociation rates obtained in the TAMU-NP framework, shown as functions of temperature at $p=0$~\cite{Wu:2025hlf}. The rates are extracted from the pole structure of the in-medium $T$-matrix in the complex energy plane, using a heavy-quark potential constrained by nonperturbative lattice-QCD Wilson line correlators. The solid lines show the inelastic dissociation rates for the $\Upsilon$(1S) (blue), $\chi_b$(1P), $\Upsilon$(2S) (orange), $\chi_b$(2P) (purple), and $\Upsilon$(3S)(red) states.}
   \label{fig:rates_nonpert}
\end{figure}

\begin{figure*}[ht]
    \centering
    \includegraphics[width=\textwidth]{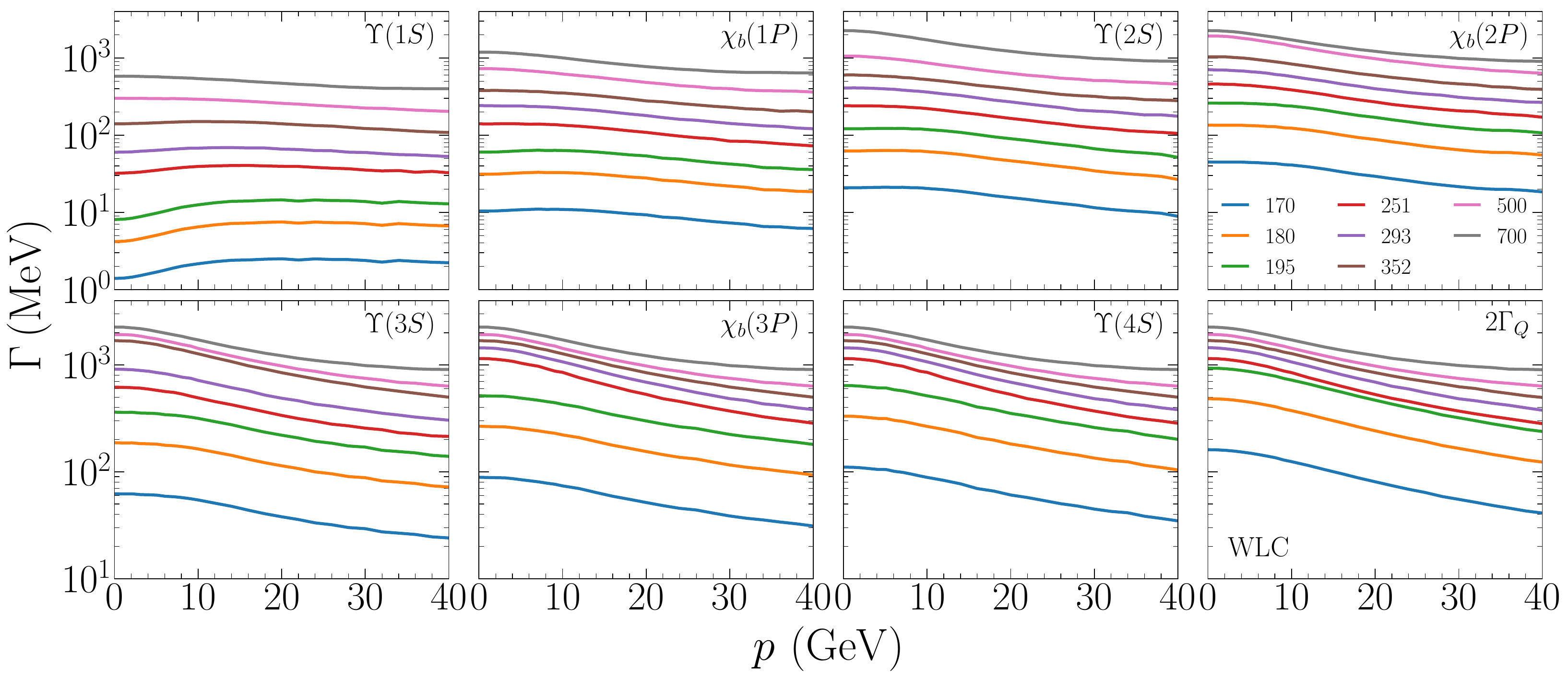}
    \caption{Nonperturbative bottomonium dissociation rates as a function of momentum over a range of temperatures~\cite{Wu:2025hlf}.  Each panel corresponds to a specific bottomonium state except the bottom right panel which shows twice the quark rates. The colors represent different temperatures, ranging from $T=170$~MeV (blue) to $T=700$~MeV (gray), as indicated in the legends. At high temperatures, the rates merge with the sum of the $b$ and $\overline b$ rates, shown in the bottom right panel.
    }
    \label{fig:rates_p}
\end{figure*}
The TAMU-NP rates are significantly larger than the TAMU-P rates, particularly for the excited states, as shown in Fig.~\ref{fig:rates_nonpert} as a function of temperature for $p=0$ and in Fig.~\ref{fig:rates_p} as a function of momentum for a range of temperatures.
These rates generally decrease with increasing momentum,
except for the $\Upsilon$(1S) rate at low temperature which is strongly suppressed at small $p$ due to the large 1S binding energy.


\subsection{3+1D anisotropic hydrodynamic QGP background}
\label{subsec:ahydro}
We model the bulk medium with 3\,+\,1D quasiparticle anisotropic hydrodynamics (aHydroQP), a far-from-equilibrium framework that evolves a nonconformal QGP with a temperature dependent quasiparticle mass matched to a realistic lattice QCD equation of state. The effective temperature is fixed by Landau matching and the microscopic relaxation time enters through a relaxation time approximation via a Boltzmann equation~\cite{Martinez:2010sc,Florkowski:2010cf,Alqahtani:2017mhy,Alalawi:2021jwn}. 

We simulate minimum-bias $p{+}\mathrm{Pb}$ collisions, for which the QGP lifetime is short ($\sim\!3$–$4$~fm/$c$). Initial conditions are set at $\tau_0=0.25$~fm/$c$ with transverse-momentum anisotropies $\alpha_{x,0}=\alpha_{y,0}=1$, longitudinal anisotropy $\alpha_{z,0}=0.2$, zero initial transverse flow, and Bjorken longitudinal flow. The longitudinal profile features a “tilted”  midrapidity plateau in spatial rapidity $\eta$ with Gaussian tails. The transverse density is a wounded nucleon plus binary collision mixture. The longitudinal width is tuned to reproduce the charged particle spectra $dN_{\rm ch}/d\eta$ at each energy. Note that it is broader at $\sqrt{s_{NN}}=8.16$~TeV than at $5.02$~TeV. The best fit central temperatures at $x{=}0, \, y{=}0, \, \eta{=}0$ and $\tau_0$ are $T_0 \simeq 480$~MeV (5.02~TeV) and $\simeq 496$~MeV (8.16~TeV). 
\begin{figure}[ht!]
\centering
\includegraphics[width=0.98\linewidth]{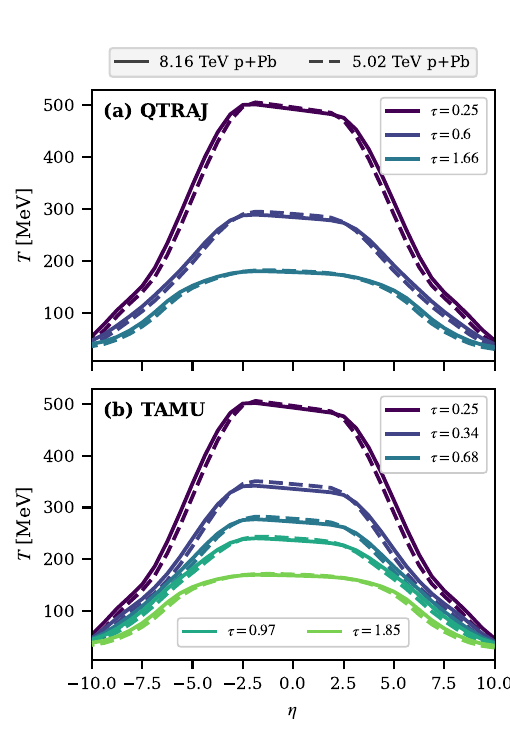}
\vspace{-3mm}
\caption{Temperature profiles $T(\eta)$ at the transverse center of the medium $(x{=}0, \, y{=}0)$ for selected proper times $\tau$ in $p{+}\mathrm{Pb}$ collisions at $\sqrt{s_{NN}}=8.16$ and $5.02$~TeV from the aHydroQP simulation. The upper (a) and lower (b) panels emphasize times $\tau$ (in fm) relevant to the QTraj and TAMU calculations, respectively.}
\label{fig:qgptemp}
\end{figure}

We employ a temperature independent shear viscosity to entropy density ratio of $\eta/s=0.32$ and determine the bulk viscosity self-consistently within the massive quasiparticle model. The background is evolved to a constant temperature freeze-out hypersurface at $T_{\rm fo}\simeq130$~MeV, after which hadrons are sampled via an extended Cooper–Frye prescription using THERMINATOR2a~\cite{Chojnacki:2011hb}. With these settings, the charged hadron observables are reproduced to within about 5\% at both energies~\cite{Strickland:2023nfm}.

Figure~\ref{fig:qgptemp} compares the evolution of the longitudinal temperature profile from the aHydroQP simulation for $p{+}\mathrm{Pb}$ collisions at $\sqrt{s_{NN}} = 5.02$ and $8.16$~TeV, for several different proper times relevant in the OQS+pNRQCD and TAMU approaches. The hydrodynamic initialization time, $\tau=0.25$~fm/$c$, is shown in both panels.  In the top panel, temperature profiles relevant for the OQS+pNRQCD calculation, labeled QTraj, are shown. The time $\tau=0.6$~fm/$c$ corresponds to when the $\Upsilon$ states begin to interact with the system in this approach.  The time at which the central temperature cools to $T_{\rm dec} = 180$~MeV, $\tau = 1.66$~fm/$c$, when the $\Upsilon$ interactions with the medium cease, is also given. In the bottom panel, the times $\tau = 0.34$, 0.68, and 0.97~fm$/c$ are representative of the bottomonium formation times in the TAMU approach, given in Table~\ref{tab:formation}.  The value $\tau = 0.34$~fm$/c$ is an average between the 1P and 2S formation times, 0.66 is the 2P formation time while 0.97 is that of the 3S. The time at which these states cease interacting with the medium at $T_{\rm dec} = 170$~MeV, $\tau = 1.85$~fm is also shown. 

The width of the profile is indicative of its energy dependence.  While the initial temperatures and temperature profiles are comparable at the two energies, the profiles for the system created at $\sqrt{s_{NN}}=8.16$~TeV are slightly broader in $\eta$ than those from $5.02$~TeV.

\subsection{Effect of primordial suppression alone}
\label{subsec:rpa_hnm}

\begin{figure*}[t!]  
\begin{center}
\includegraphics[width=0.98\linewidth]{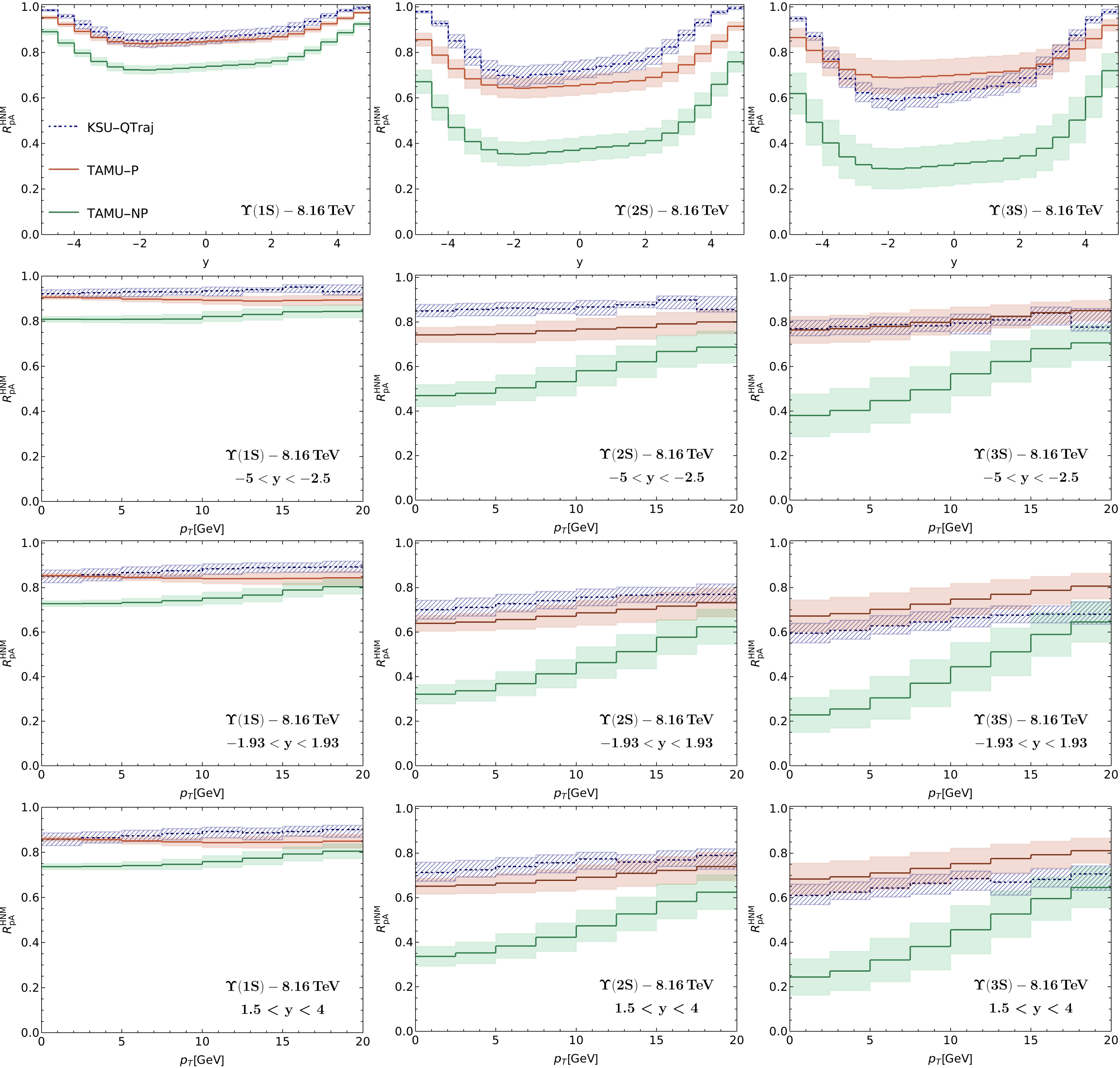}
\end{center}
\vspace{-5mm}
\caption{Hot matter suppression of $\Upsilon$(1S) (left), $\Upsilon$(2S) (middle) and $\Upsilon$(3S) (right) as a function of $y$ (top) and $p_T$ (lower rows) in backward $-5<y<-2.5$, second row from the top), central ($-1.93<y<1.93$, third row from the top) and forward ($1.5<y<4$, bottom row) rapidity regions at $\sqrt{s_{NN}} = 8.16$~TeV $p + {\rm Pb}$ collisions. The KSU-QTraj (blue band), TAMU-P (red band), and TAMU-NP (green band) results are compared. The KSU-QTraj bands are obtained by varying $\hat{\kappa} \in \{5,6,7\}$, while the uncertainties on the TAMU bands are obtained by varying the formation times.} 
\label{fig:rpA-hnm-8.16}
\end{figure*}

In this section, we present the primordial bottomonium suppression obtained using the TAMU-P and TAMU-NP rates discussed earlier.  These results are compared to those employing the KSU-Munich OQS+pNRQCD approach  (labeled KSU-QTraj in the figures). The results  at $\sqrt{s_{NN}} = 8.16$~TeV are presented as a function of $p_T$ and rapidity in Fig.~\ref{fig:rpA-hnm-8.16}.

As discussed in Ref.~\cite{Strickland:2024oat}, in the KSU-Munich approach, bottomonium is treated as an open quantum system, exchanging information with the hot QCD environment. In the hot QCD medium, the $\Upsilon$ states can change their color representation between singlet and octet until the system decouples and freezes out. The evolution of the $\Upsilon$ through the surrounding environment is described by the Lindblad master equation which depends on the $\Upsilon$ Hamiltonian and jump operators describing how the states transition between singlets and octets.  The Hamiltonian includes singlet and octet components with real and imaginary parts.  The imaginary parts describe the modification of the bottomonium widths in the medium and depend on the transport coefficients, \ie, the momentum diffusion coefficient $\kappa$ and the dispersive counterpart $\gamma$ defined in terms of the chromo-electric field correlators (see eqs. (A.1) and (A.2) in Ref.~\cite{Brambilla:2022ynh}). We vary the scaled dimensionless transport coefficients $\hat{\kappa} = \kappa/T^3$ and $\hat{\gamma} = \gamma/T^3$ in the appropriate range to obtain the theoretical uncertainty on the model. The heavy quark diffusion coefficient $\hat{\kappa}$ is varied in the range $\hat{\kappa} = 6 \pm 1$ as predicted from direct and indirect lattice calculations. The central value of $\hat{\kappa}$ is appropriate for conditions at the LHC.  The dispersion coefficient $\hat{\gamma}$, responsible for the thermal mass shift in medium, is assumed to be zero for the bottomonium states, consistent with recent lQCD results~\cite{Larsen:2019bwy}. The Lindblad equation, including both real and imaginary parts, is solved in the evolving hydrodynamic background. The modifications of the widths of the different bottomonium states in the hot medium result in significantly different suppression patterns, cf.~Ref.~\cite{Strickland:2024oat} for full details.

As described in Sec.~\ref{subsec:ahydro}, the bottomonium evolution in this approach starts from the vacuum potential at initial time $\tau_0 = 0.25$~fm/$c$ while the medium effects are assumed to set in at $\tau_{\rm med} \sim 0.6$~fm/$c$ and continue evolving until the decoupling temperature, $T_{\rm dec} = 180$~MeV in Ref.~\cite{Strickland:2024oat}. This value is not adjusted here.  However, we note that in the TAMU approach, the system is evolved to $T_{\rm dec} = 170$~MeV~\cite{Du:2017qkv}.  This value is used in the TAMU-P and TAMU-NP calculations.  The difference in the decoupling temperatures has a very small effect on $R_{pA}$.  If the TAMU evolution was halted at 180~MeV, the suppression would be reduced by 1-2\% for TAMU-P and 2-3\% for TAMU-NP.  The reduction in suppression is larger for TAMU-NP because the overall rates are larger.

The initial proto-bottomonium production mechanism is through the CEM, as explained in Sec.~\ref{sec:cnm}, for all HNM calculations.  However, the bottomonium formation process is different in the KSU-Munich and TAMU approaches.
In the KSU-Munich approach, the bottomonia states are assumed to be produced instantaneously, without any formation time. In the TAMU approach, formation times based on the bottomonium binding energies are employed, proportional to $1/E_B$ with a coefficient varied between 1 and 2. 
These times adhere to the concept that the $b \overline b$ pair must resolve the energy gap between the bound state and the continuum which sets the timescale for formation~\cite{Song:2013}. Loosely bound states with small $E_B$, such as $\Upsilon$(3S), take longer to form while tightly bound states such as $\Upsilon$(1S) form quickly. 
The same formation times are used with the TAMU-P and TAMU-NP rates here, based on the vacuum binding energies listed in Table~\ref{tab:formation}. 
We note that in the previous work~\cite{Du:2017qkv}, where the TAMU-P rates were originally used, the $\Upsilon$(1S), $\Upsilon$(2S), and $\Upsilon$(3S) formation times were taken to be 0.5~fm/$c$, 1~fm/$c$, and 1.5~fm/$c$, respectively, leading to somewhat lower suppression.
\begin{table}[htb]
\centering
\begin{tabular}{|c|c|c|}
\hline
State & $E_B$ (GeV) & \(\bm{\tau_{\rm form}}\) $\sim 1/E_B$ (fm/\(c\)) \\ 
\hline
1S & 1.10 & 0.18 \\ 
\hline
1P & 0.66 & 0.30 \\ 
\hline
2S & 0.54 & 0.37 \\ 
\hline
2P & 0.30 & 0.66 \\ 
\hline
3S & 0.20 & 0.97 \\ 
\hline
\end{tabular}
\caption{Vacuum binding energies \(E_B\)~\cite{ZhanduoTang:2023pla} and pertinent formation times \(\tau_{\rm form}\) for bottomonium states used in the TAMU approaches.}
\label{tab:formation}
\end{table}

Figure~\ref{fig:rpA-hnm-8.16} shows the HNM modification factor due to primordial suppression, $R_{pA}^{\rm HNM}$, as a function of rapidity (top row) for the $\Upsilon$(1S) (left), $\Upsilon$(2S) (middle) and $\Upsilon$(3S) (right) states in $p+{\rm Pb}$ collisions at 8.16~TeV for all three reaction rates studied. The rapidity dependence has a characteristic shape, with little suppression at forward and backward rapidities, $|y| > 2.5$, where there is less hot matter created. Near midrapidity, the suppression is largest at backward rapidity, $y \sim -2$, in the lead-going direction where the hot matter density is greatest. 

This asymmetric behavior reflects the shape of the final-state light hadron multiplicity distributions calculated with aHydro~\cite{Strickland:2024oat}. The KSU-QTraj result has a slightly different shape as a function of rapidity, with less suppression at forward and backward rapidity and somewhat stronger suppression in the lead-going direction at $y \sim -2$.

There is a consistent increase in suppression as the $\Upsilon$ mass increases and the binding energy decreases. The KSU-QTraj (light blue) and TAMU-P (dark red) results are in relatively good agreement for all three states although the KSU-QTraj suppression tends to be somewhat weaker.  The suppression is significantly stronger for the TAMU-NP reaction rates, suggesting that the nonperturbative calculation introduces effects not accounted for in either the KSU-QTraj or the TAMU-P approaches.  The difference in suppression grows larger as the $\Upsilon$ states become more weakly bound, indicating a greater sensitivity of the excited states to nonperturbative effects.

Figure~\ref{fig:rpA-hnm-8.16} also shows the HNM  effects as a function of $p_T$ (second, third and fourth rows) for backward rapidity ($- 5 < y < -2.5$, left column), midrapidity ($-1.93 < y < 1.93$, middle column) and forward rapidity ($1.5 < y < 4$, right column).  The $\Upsilon$(1S), $\Upsilon$(2S) and $\Upsilon$(3S) results are shown in the first, second and third columns, respectively.  The same trends seen as a function of rapidity are also seen here, with the KSU-QTraj and TAMU-P results generally similar and considerably stronger suppression for the TAMU-NP rates. Both of the TAMU rates exhibit a somewhat stronger $p_T$ dependence, with greater suppression at low $p_T$ and decreasing suppression at higher $p_T$.

The suppression as a function of $p_T$ is strongest at midrapidity and forward rapidity, likely because these regions overlap from $1.5 < y < 1.93$. It is weakest at backward rapidity, $-5 < y < -2.5$, even though the rapidity-dependent suppression is strongest in the lead-going direction.  This is because, even though the suppression is maximized at $y \sim -2.5$, it decreases quickly at more negative rapidities, as shown in the top row of Fig.~\ref{fig:rpA-hnm-8.16}. The same decreasing suppression also applies to forward rapidity.  However, in this case, the rapidity range, $1.5 < y < 4$, includes more of the region of stronger suppression than at backward rapidity.  The results are similar for the KSU-QTraj and TAMU-P results for the three states as a function of $p_T$ over most of the rapidity range. The TAMU-NP rates produce the greatest suppression as well as the strongest $p_T$ dependence.

We have also computed the hot matter suppression at $\sqrt{s_{NN}} = 5.02$~TeV. At this lower energy, the suppression is slightly weaker, as expected due to the lower temperature. The energy dependence predicted by the KSU-QTraj calculation is somewhat stronger than that obtained with the TAMU rates.

Our calculations discussed thus far do not yet include  regeneration. The previous KSU-QTraj calculations found rather good agreement with the data without regeneration, indicating that the need for this effect in $p+{\rm Pb}$ collisions could be small. Previous semiclassical simulations~\cite{Du:2017qkv}, where the TAMU-P rates were used, predicted some regeneration in $A+A$ collisions that becomes more significant, or even dominant, for the strongly suppressed excited states in more central collisions. On the other hand, the results with the TAMU-NP rates suggest that substantial regeneration is required to produce agreement with the data~\cite{Wu:2025lcj}. 
We test the relevance of regeneration for bottomonium in Sec.~\ref{sec:reg} where we calculate its magnitude and include it in our comparison with the data.

\subsection{Feeddown contribution}
\label{subsec:feeddown}
As the bottomonium states propagate out of the hot QGP, the excited states can decay to lower-lying states, feeding these channels.  To account for feeddown, we define a matrix ($F$), which empirically relates the experimentally observed cross sections ($\vec{\sigma}_{\text{exp}}$) to the directly produced cross sections ($\vec{\sigma}_{\text{direct}}$) in $p+p$ collisions, \mbox{$\vec{\sigma}_{\text{exp}} = F \vec{\sigma}_{\text{direct}}$}. The quantities $\vec{\sigma}_{\text{direct}}$ and $\vec{\sigma}_{\text{exp}}$ contain the production cross sections for the bottomonium states $\Upsilon$(1S), $\Upsilon$(2S), $\chi_{b0}$(1P), $\chi_{b1}$(1P), $\chi_{b2}$(1P), $\Upsilon$(3S), $\chi_{b0}$(2P), $\chi_{b1}$(2P), and $\chi_{b2}$(2P) states before and after feeddown, respectively. Further details on calculating $F$ are provided in Refs.~\cite{Boyd:2023ybk,Strickland:2024oat}. Incorporating feeddown contributions,  
the nuclear modification factor $R_{\rm pA}$ in min-bias $p+{\rm Pb}$ collisions for a given bottomonium state $i$ 
is expressed as:
\beq
R^{\Upsilon, \rm incl}_{pA}(p_T,y) = \frac{\left(F \cdot R^{\Upsilon \, \rm direct}_{pA}(p_T,y) \cdot \vec{\sigma}_{\text{direct}}\right)^{i}}{\vec{\sigma}_{\text{exp}}^{i}} \, ,
\label{eq:feeddown}
\eeq
where $R^{\Upsilon \, \rm direct}_{pA}(p_T,y)$ is the nuclear modification factor before the feeddown contributions. 

The experimentally measured bottomonium cross sections are $d \vec{\sigma}_{\text{exp}}/dy = \{57.6$, 19, 3.72, 13.69, 16.1, 6.8, 3.27, 12.0, $14.15\}$ nb for $\sqrt{s_{NN}} = 5.02$~TeV $p+p$ collisions, as reported in Refs.~\cite{Strickland:2023nfm, Aaij:2014caa, Brambilla:2020qwo} in the rapidity interval $\Delta y = 1.8$.  To obtain the corresponding cross sections at $\sqrt{s_{NN}} =  8.16$~TeV, $\vec \sigma_{\rm exp}$ is scaled by the square-root of the ratio of the center-of-mass energies, \mbox{$\sqrt{8.16/5.02} = 1.275$}. 

\subsection{CNM and Primordial Suppression}
\label{subsec:results_wout_reg}

We now present the combined results of primordial suppression with the total cold nuclear matter effects calculated in Sec.~\ref{sec:cnm}. 
The total nuclear modification factor, including both the CNM and HNM effects on the inclusive $R^\Upsilon_{pA}$, defined in Eq.~(\ref{eq:feeddown}), is
\beq
R^{\Upsilon}_{pA} = R^{\rm CNM}_{pA}  \times  R^{\rm HNM}_{pA} \, ,
\label{eq:rpa_nofeeddown}
\eeq
where the suppression due to CNM alone is given in Eq.~(\ref{eq:rpa_CNM}). The factor $R^{\rm HNM}_{pA}$ is the primordial suppression factor due to the hot QGP medium calculated earlier in this section.  We note that because the CNM effects are independent of the final bottomonium state in our approach, feeddown only affects the primordial suppression in Eq.~(\ref{eq:rpa_nofeeddown}).

We present the nuclear suppression factors for the bottomonium states, including feeddown, based on Eq.~(\ref{eq:feeddown}). Hot matter effects are shown for all three cases described here: KSU-QTraj, TAMU-P and TAMU-NP.  We first discuss $\Upsilon$($n$S) suppression in $p +{\rm Pb}$ collisions at $\sqrt{s_{NN}} = 5.02$ and 8.16~TeV. Our results are compared with experimental data from ALICE~\cite{ALICE:2019qie}, ATLAS~\cite{ATLAS:2017prf}, CMS~\cite{CMS:2022wfi}, and LHCb~\cite{LHCb:2018psc}.  In addition, we present predictions for suppression of the $\chi_b$($n$P) states at 8.16~TeV. 

The uncertainties of $R^{\Upsilon}_{pA}$ in Eq.~(\ref{eq:rpa_nofeeddown}), $\Delta R^{\Upsilon}_{pA}$, are obtained by adding the uncertainties of 
both $R^{\rm CNM}_{pA}$
and $R^{\rm HNM}_{pA}$ in quadrature,
\begin{equation}
    \begin{aligned}
        R^{\Upsilon}_{\rm pA}&=\left(R^{\rm CNM}_{pA}\pm \Delta R^{\rm CNM}_{pA}\right)\left(R^{\rm HNM}_{pA}\pm\Delta R^{\rm HNM}_{pA}\right)\\
        &\simeq R^{\rm CNM}_{pA} R^{\rm HNM}_{pA} \pm \Delta R^{\rm CNM}_{pA} R^{\rm HNM}_{pA}\pm R^{\rm CNM}_{pA}\Delta R^{\rm HNM}_{pA}\\
        &= \left(R^{\rm CNM}_{pA} R^{\rm HNM}_{pA} \pm \Delta R^{\Upsilon}_{\rm pA}\right)\,,
    \end{aligned}
\end{equation}
where $\Delta R^{\Upsilon}_{pA}$ is determined by the expression
\begin{equation}
\Delta R^{\Upsilon}_{pA}= R^{\Upsilon}_{pA}\sqrt{\left(\frac{\Delta R^{\rm CNM}_{pA}}{R^{\rm CNM}_{pA}}\right)^2 + \left(\frac{\Delta R^{\rm HNM}_{pA}}{R^{\rm HNM}_{pA}}\right)^2} \ .
\label{eq:uncertainty}
\end{equation}
As noted in Sec.~\ref{sec:cnm}, the only CNM uncertainty taken into account is that of the nPDFs.  There are no parameters varied in the calculation of energy loss and momentum broadening.  The variation in $R_{pA}^{\rm HNM}$ comes from $\hat \kappa$ in the KSU-QTraj approach and the formation times in the TAMU approaches.

The differences in suppression among the bottomonia states arise entirely from the HNM effects, which depend on the respective masses, binding energies, formation times, and dissociation rates of the states. The effect of CNM energy loss, as implemented in our framework, employs an average mass for all the bottomonium states. The nPDF effects depend on the $b \overline b$ pair mass and not the final-state mass.  In the figures, the CNM effects alone are shown by the gray bands while the results that include both the CNM and hot matter effects are denoted as in Figs.~\ref{fig:rpA-hnm-8.16}, with KSU-QTraj in blue, TAMU-P in red and TAMU-NP in green.  The total uncertainties for the CNM + HNM effects include the nPDF uncertainties and the variation of the diffusion coefficient $\hat{\kappa}$ (KSU-QTraj) and formation times (TAMU-P and TAMU-NP).  The CNM and HNM uncertainties are added in quadrature to obtain the full uncertainty bands for each state.
\begin{table}[h]
\begin{tabular}{|c|c|c|}
\hline
Name & Mass (GeV) & Degeneracy \\
\hline
$\eta_b$(1S) & 9.343 & 1 \\
$\Upsilon$(1S) & 9.4603 & 3 \\
$\chi_{b0}$(1P) & 9.8594 & 1 \\
$\chi_{b1}$(1P) & 9.8928 & 3 \\
$\chi_{b2}$(1P) & 9.9122 & 5 \\
$\Upsilon$(2S) & 10.023 & 3 \\
$\chi_{b0}$(2P) & 10.233 & 1 \\
$\chi_{b1}$(2P) & 10.255 & 3 \\
$\chi_{b2}$(2P) & 10.269 & 5 \\
$\Upsilon($3S) & 10.3552 & 3 \\
$\Upsilon$(4S) & 10.5794 & 3 \\
$\Upsilon(10860)$ & 10.865 & 3 \\
$\Upsilon(11020)$ & 11.019 & 3 \\
\hline
\end{tabular}
\caption{Masses and degeneracies of the bottomonium states~\cite{ParticleDataGroup:2024cfk}  \label{Ups_degen_table}.}
\end{table}

The TAMU calculations employ the masses given in Table~\ref{Ups_degen_table}.
Spin-weighted averages are used for the $\chi_b$ states:
9.8881~GeV and 10.2523~GeV for the $\chi_{b1}$(1P) and $\chi_{b2}$(2P) states respectively.  Recall that there is no distinction between $\Upsilon$(3S) and $\chi_b$(2P) states for TAMU-P while the two are treated separately in the TAMU-NP calculations, compare Figs.~\ref{fig:rates_pert} and \ref{fig:rates_p}.

\subsubsection{$\Upsilon$($n$S) Suppression}
\label{subsubsec:upsilon}
In this section, we present the final results for the primordial $R^{\Upsilon(n{\rm S})}_{pA}$ as a function of rapidity and $p_T$ for the $\Upsilon(n{\rm S})$ states at $\sqrt{s_{NN}} = 5.02$ and 8.16~TeV in $p + {\rm Pb}$ collisions.

\begin{figure*}[ht!]
\begin{center}
\includegraphics[width=1.0\linewidth]{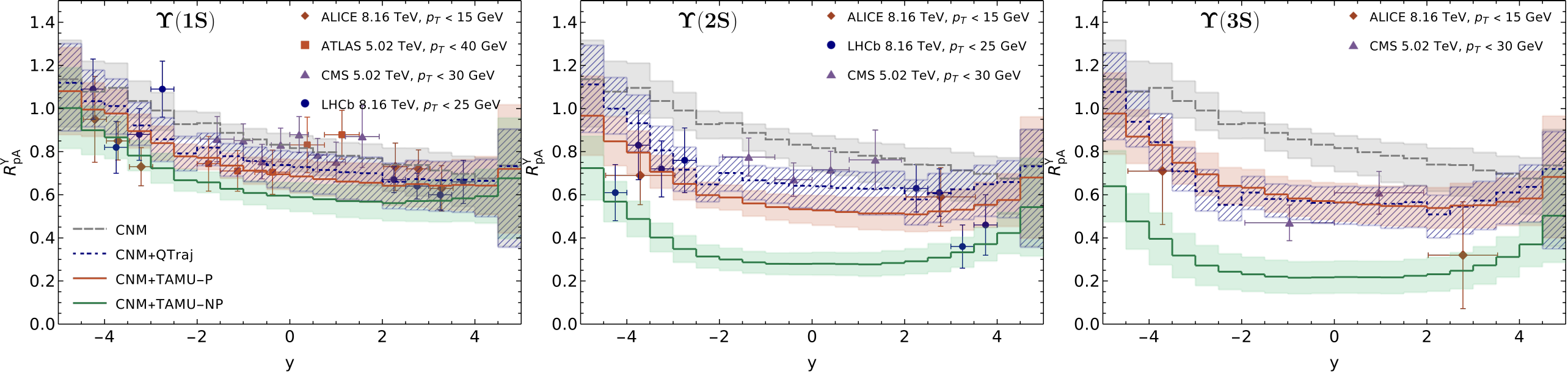}
\end{center}
\vspace{-5mm}
\caption{The suppression factor, including both cold and hot nuclear matter, for $\Upsilon$(1S) (left), $\Upsilon$(2S) (middle) and $\Upsilon$(3S) (right) as a function of rapidity. The results for CNM alone (gray band), CNM+KSU-QTraj (blue band), CNM+TAMU-P (perturbative rates, dark red band), and CNM+TAMU-NP (nonperturbative rates, light green band) are compared. 
At $|y| < 2$ and $|y| \ge 2$, the results for $\sqrt{s_{NN}} =  5.02$~TeV and $\sqrt{s_{NN}} = 8.16$~TeV $p+{\rm Pb}$ are shown respectively. The calculations are compared to data from ALICE~\cite{ALICE:2019qie}, ATLAS~\cite{ATLAS:2017prf}, CMS~\cite{CMS:2022wfi}, and LHCb~\cite{LHCb:2018psc}.
} 
\label{fig:rpA-y-noReg}
\end{figure*}

Figure~\ref{fig:rpA-y-noReg} shows the results as a function of rapidity compared to data from ALICE~\cite{ALICE:2019qie}, ATLAS~\cite{ATLAS:2017prf}, CMS~\cite{CMS:2022wfi}, and LHCb~\cite{LHCb:2018psc}.   The results are given for $\sqrt{s_{NN}} = 5.02$~TeV for central rapidity, $|y| < 2$, and for $\sqrt{s_{NN}} = 8.16$~TeV at forward and backward rapidity, $|y| \ge 2$, allowing for direct comparison with the corresponding experimental data. The results are shown for $\Upsilon$(1S) (left), $\Upsilon$(2S) (center), and $\Upsilon$(3S) (right). The data are compared to CNM effects alone (gray band), CNM+KSU-QTraj (blue band), CNM+TAMU-P (red band), and CNM+TAMU-NP (green band). 

In the case of $\Upsilon$(1S), adding hot matter suppression has only a small effect on the midrapidity results.  The addition of HNM effects to the CNM calculations produces only a modest increase in suppression, giving a slightly greater total suppression, particularly at midrapidity.  All the calculations lie within the experimental uncertainties, showing that the HNM effects do not add much suppression beyond those of cold matter. 

However, the CNM suppression is insufficient to describe the $\Upsilon$(2S) and $\Upsilon$(3S) data, as was found in our previous work with KSU-QTraj alone~\cite{Strickland:2024oat}.    Note that the KSU-QTraj and TAMU-P results agree well with the data at midrapidity for the excited $\Upsilon$ states and almost entirely overlap, with small differences visible for the $\Upsilon$(2S).  On the other hand, the new TAMU-NP rates predict significantly more suppression than the midrapidity data allow.  We note that at the most forward and backward rapidities the TAMU-NP rates give somewhat better agreement with the 8.16~TeV data.

\begin{figure*}[ht!]
\begin{center}
\includegraphics[width=0.98\linewidth]{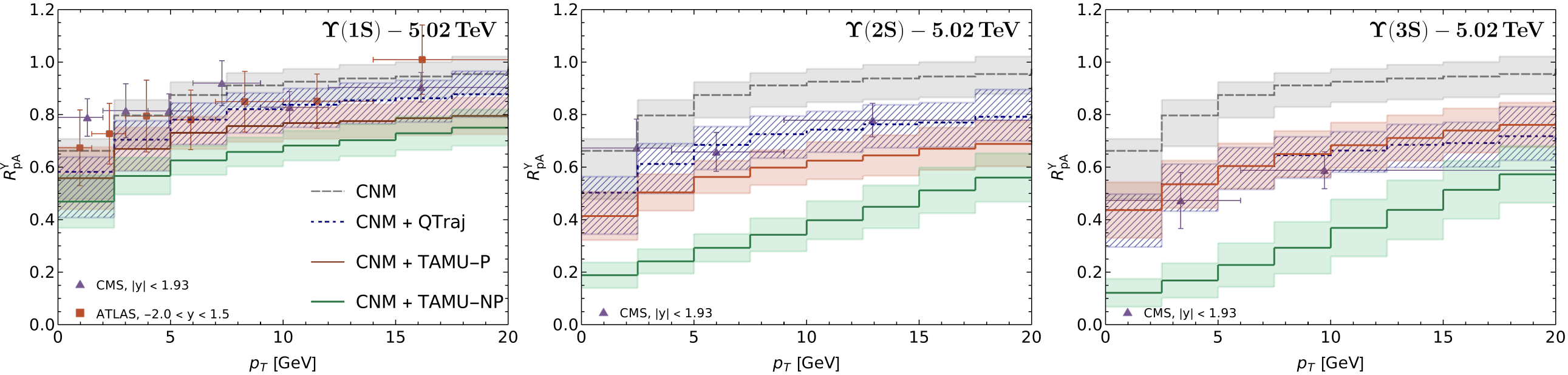}
\vspace{-5mm}
\end{center}
\caption {The nuclear suppression factor, including both cold and hot nuclear matter, for $\Upsilon$(1S) (left), $\Upsilon$(2S) (middle) and $\Upsilon$(3S) (right) as a function of $p_T$ at $\sqrt{s_{NN}} = 5.02$~TeV. The results are given for CNM alone (gray band), CNM+KSU-QTraj (blue band), CNM+TAMU-P (red band), and CNM+TAMU-NP (green band). 
The calculations are compared to data from ATLAS~\cite{ATLAS:2017prf} and CMS~\cite{CMS:2022wfi}. The horizontal uncertainties indicate the width of the reported $p_T$ bins.
}
\label{fig:rpA-pt-midRap-5.02}
\end{figure*}

Figure~\ref{fig:rpA-pt-midRap-5.02} presents the results as a function of $p_T$ at midrapidity for $\sqrt{s_{NN}} = 5.02$~TeV.   The trends are similar to those seen as a function of rapidity in Fig.~\ref{fig:rpA-y-noReg}.  All the calculations are generally compatible with the $\Upsilon$(1S) data while hot nuclear matter effects are needed to describe the excited $\Upsilon$ states.  The CNM+KSU-QTraj result for $\Upsilon$(2S) shows slightly less suppression than the CNM+TAMU-P result while, for the $\Upsilon$(3S), the two calculations overlap.  Both agree with the excited state data.  However, the results for CNM+TAMU-NP show considerably more suppression than the data. 

\begin{figure*}[hb!]
\begin{center}
\includegraphics[width=0.980\linewidth]{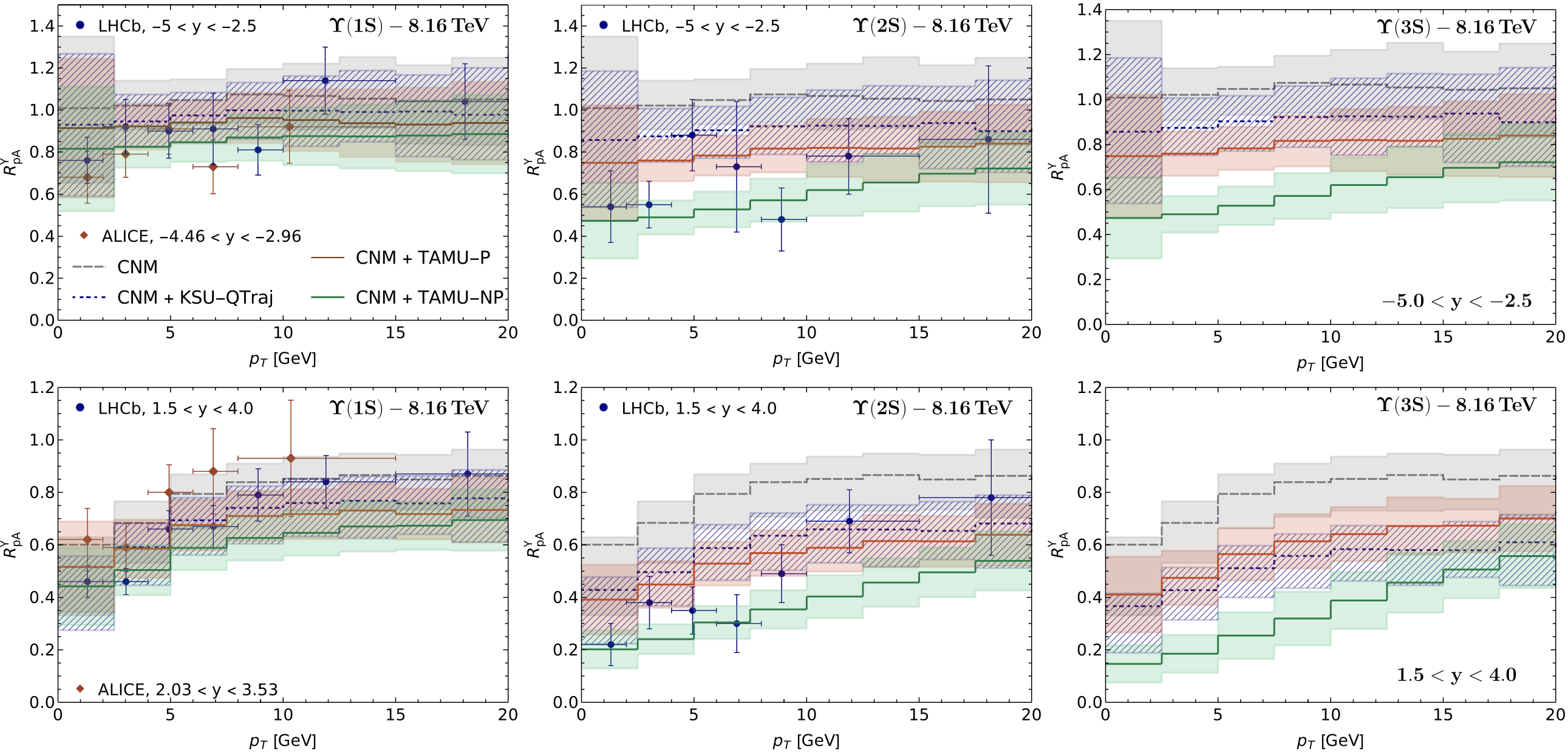}
\vspace{-5mm}
\end{center}
\caption{The nuclear suppression factor, including both cold and hot nuclear matter, for $\Upsilon$(1S) (left) and $\Upsilon$(2S) (right) as a function of $p_T$ at $\sqrt{s_{NN}} = 8.16$~TeV. The CNM alone (gray band), CNM+KSU-QTraj (blue band), CNM+TAMU-P (red band), and CNM+TAMU-NP (green band) results are compared. 
The top row shows results at backward rapidity while the bottom row presents the results at forward rapidity.
The calculations are compared to data from ALICE~\cite{ALICE:2019qie} and LHCb~\cite{LHCb:2018psc}.
The horizontal uncertainties indicate the width of the reported $p_T$ bins.
}  
\label{fig:rpA-pt-farRap-8.16}
\end{figure*}

Figure~\ref{fig:rpA-pt-farRap-8.16} shows $R^{\Upsilon(nS)}_{pA}$ as a function of $p_T$ at backward (top row) and forward (bottom row) rapidity in 8.16~TeV $p + {\rm Pb}$ collisions.  At backward rapidity, the large uncertainty band in the lowest $p_T$ interval is due to CNM shadowing alone.  The trends are generally the same as seen at midrapidity for 5.02~TeV although the separation between the calculations for the excited states is somewhat reduced relative to the midrapidity region shown previously for 5.02~TeV. 

\subsubsection{$\chi_b$ Predictions}
\label{subsubsec:chistates}
\begin{figure*}[ht!]
\begin{center}
\includegraphics[width=0.98\linewidth]{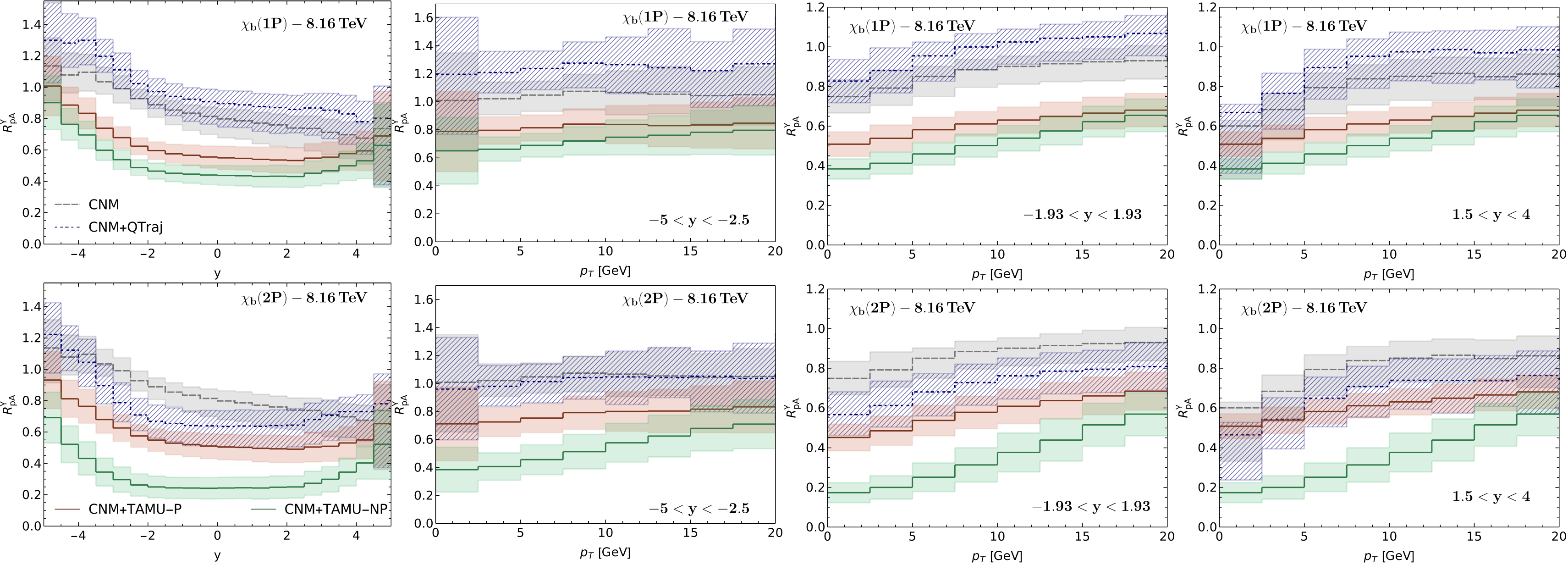}
\vspace{-5mm}
\end{center}
\caption{The nuclear suppression factor $R_{pA}$ for $\chi_b$(1P)  (top row) and $\chi_b$(2P) (bottom row) as a function of $y$ (left-most column), and $p_T$ in backward rapidity (center left), central rapidity (center right) and forward rapidity (right-most column) in $\sqrt{s_{NN}} = 8.16$~TeV $p + {\rm Pb}$ collisions. The results are given for CNM effects alone (gray band), CNM+QTraj (blue band), CNM + TAMU-P (red band) and CNM+TAMU-NP (green band).}  
\label{fig:rpA-xb-8TeV}
\end{figure*}

In addition to considering $\Upsilon$($n$S) suppression, in this section we present predictions of the nuclear modifications of the $\chi_b$ states at $\sqrt{s_{NN}} = 8.16$~TeV.  The effects will be similar albeit somewhat reduced for 5.02~TeV.  The results are presented in Fig.~\ref{fig:rpA-xb-8TeV}. 

The cold nuclear matter effects are identical to those of the $\Upsilon$($n$S) states. These effects are independent of the angular momentum or degeneracy of the states.

In most cases shown here, the CNM effects are the baseline for the minimal suppression effect.  However, in the KSU–Munich framework, OQS+pNRQCD, there is a pronounced {\em enhancement} of the $\chi_b$(1P) in $p+\mathrm{Pb}$ collisions at $\sqrt{s_{NN}}=8.16\,$~TeV, as can be seen in Fig.~\ref{fig:rpA-xb-8TeV}. There is a slight corresponding enhancement of the $\chi_b$(2P) over the CNM effects alone at the most forward and backward rapidities but there is suppression at midrapidity.  The counterintuitive $\chi_b$(1P) enhancement is a feature of the approach and is independent of whether or not quantum-jump channels, responsible for quantum regeneration effects~\cite{Brambilla:2023hkw}, are included. This P-wave enhancement suggests the potential for quantum state mixing due to the time-dependent in-medium potential~\cite{Islam:2020gdv}. As the QGP expands and cools, the real part of the heavy quark potential evolves, inducing non-adiabatic couplings between instantaneous eigenstates of the Hamiltonian and causing quantum state mixing. This effect, not present in semiclassical approaches, is a unique prediction of the quantum transport of quarkonium in the evolving hot medium.

In contrast, the TAMU-P and TAMU-NP treatments both predict $\chi_b$ suppression. As shown for the TAMU-P rates in  Fig.~\ref{fig:rates_pert}, the $\chi_b$(1P) rate is much larger than that of the $\Upsilon$(1S) in the low temperatures range relevant in $p+{\rm Pb}$ collisions. Thus one can discern a separation between the $\chi_b$(1P) predictions at midrapidity that is absent for the $\Upsilon$(1S) in Fig.~\ref{fig:rpA-y-noReg}.  However, as noted earlier, the TAMU-P treatment did not predict separate rates for the $\chi_b$(2P) and $\Upsilon$(3S) but treated them equivalently.  The only differences between these states here then are due to different feeddown contributions.  The TAMU-NP treatment includes different rates for the $\Upsilon$(3S) and $\chi_b$(2P), as shown in Fig.~\ref{fig:rates_p}. 

The difference between the two TAMU approaches to $\chi_b$ suppression are apparent in Fig.~\ref{fig:rpA-xb-8TeV}, both as a function of rapidity and as a function of $p_T$.  There is less difference between the suppression in the TAMU-P and TAMU-NP rates for the $\chi_b$(1P) than for the $\chi_b$(2P).

\section{Quarkonium regeneration in $p+\rm Pb$ Collisions}
\label{sec:reg}
In this section, we discuss our evaluation of bottomonium regeneration in $p+$Pb collisions, as dictated by the principle of detailed balance in kinetic approaches. This kinetic approach is coupled to an instantaneous coalescence model to calculate the regenerated $p_T$ spectra. One may distinguish two ways of regenerating bottomonium states in a heavy-ion collision. First, even though the initially produced quarkonium states are suppressed, the correlation between the dissociated $b\overline b$ pair remains, making it possible for them to find each other and bind again after their initial dissociation.
Second, uncorrelated $b$ and $\overline b$ quarks can form a pair and bind, creating new quarkonium states. However, except for very central Pb+Pb collisions at the LHC, no more than one $b\overline b$ pair is present per unit of rapidity. Thus this second case is irrelevant for $p+A$ collisions. Note, however, that in a $p+p$ collision less than 0.5\% of all produced $b\overline b$ pairs forms a bottomonium state. Thus, even if only 1 in 200 $b\overline b$ pairs leads to bottomonium  formation in $p+A$ (or $A+A$) collisions, its contribution to $R_{pA}$ is of order one. Thus, regeneration of bottomonium cannot a priori be considered as negligible, even if the number of primordially produced $b \overline b$ pairs is small. The number of primordial pairs is defined as
\beq
N_{b \overline b}^{\text{prim}}  = N_{\text{coll}} \frac{\sigma^{\text{tot}}_{pN \xrightarrow{} b \overline b}}{\sigma^{\text{in}}_{pN}} \,  , 
\label{eq:N_bbbar_pp} 
\eeq
and is a key ingredient in the computation of the regenerated yields. Here, $\sigma^{\text{tot}}_{pN \xrightarrow{} b \overline b} = \Delta y \times 34.5 \, \mu b$ with $\Delta y = 1.8$ at $\sqrt{s_{\rm NN}} = 5.02$~TeV~\cite{ALICE:2021mgk}.
These regenerated yields can be extracted from the rate equation by subtracting the primordial component defined in Eq.~\eqref{eq:rate-eq-prim} from the total rate in Eq.~\eqref{eq:rate-eq} to obtain
\beq
\frac{d N_{\Upsilon}^{\text{reg}} (\tau)}{d \tau} = - \Gamma_{\Upsilon} (T(\tau)) [N_{\Upsilon}^{\text{reg}} (\tau) - N_{\Upsilon}^{\text{eq}} (\tau)] \, ,
\label{eq:rate-eq-reg}
\eeq
which can be rearranged as
\beq
\frac{d N_{\Upsilon}^{\text{reg}} (\tau)}{d \tau} + \Gamma_{\Upsilon} (T(\tau)) N_{\Upsilon}^{\text{reg}} (\tau) =  \Gamma_{\Upsilon} (T(\tau))  N_{\Upsilon}^{\text{eq}} (\tau) \, .
\label{eq:rate-eq-reg2}
\eeq
Equation~\eqref{eq:rate-eq-reg2} can be solved by multiplying by the factor $\mu(\tau) = \exp(\int \Gamma_\Upsilon (\tau) d\tau)$ and integrating over $\tau$.  The lower limit on the $\tau$ integration is the dissociation time, $\tau_{\text{diss}}$, \ie, the time the cooling fireball reaches the dissociation temperature $T_{\text{diss}}$, of a pertinent state.  Regeneration is possible at temperatures lower than $T_{\rm diss}$ ($N^{\text{reg}}_{\Upsilon} (\tau_{\text{diss}}) = 0$).  The integration ends when the system freezes out at $\tau = \tau_f$ , where the final temperature $T_f$ is the decoupling temperature $T_{\rm dec} = 170$~MeV.  Dividing both sides by $\mu(\tau)$, we have
\bqa
N_{\Upsilon}^{\text{reg}} (\tau) &&= \exp \left(- \int_{\tau_{\text{diss}}}^{\tau} \Gamma_{\Upsilon} (T(\tau'')) d\tau'' \right ) \nonumber \\ 
 && \!\!\!\!\!\!\!\!\!\!\!\!\!\!\!\!\! \times \int_{\tau_{\text{diss}}}^{\tau} \Gamma_{\Upsilon} (T(\tau')) N_\Upsilon^{\text{eq}}(\tau') \exp \left(\int_{\tau_{\text{diss}}}^{\tau'} \Gamma_{\Upsilon} (T(\tau'')) d\tau'' \right) d\tau' \nonumber \\
&&\!\!\!\!\!\!\!\!\!\!\!\!\!\!\!\!\! = \int_{\tau_{\text{diss}}}^{\tau} \Gamma_{\Upsilon} (T(\tau')) N_\Upsilon^{\text{eq}}(\tau') \exp \left(- \int_{\tau'}^{\tau} \Gamma_{\Upsilon} (\tau'') d\tau'' \right)  d\tau' \, .\nonumber \\  
\label{eq:rate-eq-reg-sol}
\eqa
The factor $\exp(- \int_{\tau'}^{\tau} \Gamma_{\Upsilon} (\tau'') d\tau'' )$ allows for suppression of the newly regenerated bottomonium states.
To solve this rate equation, we employ the same hydrodynamic evolution, aHydroQ, as described above along the same trajectories as used in the suppression calculations.

In the remainder of this section, we first determine the required additional transport parameter, the equilibrium limit for each state, $N_\Upsilon^{\rm eq}$, in Sec.~\ref{ssec:Neq}; then describe the coalescence of $b$ and $\overline b$ quarks to form bottomonium and compute their $p_T$ spectra in Sec.~\ref{ssec:coal}; and finally determine the overall final nuclear modification factor with regeneration included for the $\Upsilon$($n$S) and $\chi_b$($n$P) states in Sec.~\ref{ssec:Rpa-regen}. 

We include regeneration only for the TAMU-NP calculations which exhibit the largest primordial suppression.  The lower TAMU-P rates would have a correspondingly smaller contribution from regeneration. Quantum regeneration can also be included in the KSU-QTraj approach by turning on quantum jumps. These have been turned off in the current calculations but we have verified that, for min-bias $p + {\rm Pb}$ collisions, there is less than a 5\% effect on the final yields. The jump operators can lead to regeneration from correlated $b\overline b$ pairs that form bottomonium states in $p+p$ collisions.  However, we stress that this does not include contributions from the large number of $b\overline b$ configurations that do not form bottomonia in $p+p$ collisions, as mentioned above. 

\subsection{Equilibrium limit}
\label{ssec:Neq}
The equilibrium limit of bottomonium production, $N^{\rm eq}_{\Upsilon}$, necessary for calculating $\Upsilon$ regeneration, is determined by assuming relative chemical equilibrium between the open and hidden bottom (bottomonium)  states at a given time and volume of the system, keeping the total number of $b\overline b$ pairs conserved throughout the expansion of the fireball~\cite{Emerick:2011xu, Du:2017qkv}. The equilibrium number of $\Upsilon$ states can be written in terms of the thermal densities of the bottomonium states $n_\Upsilon$ as~\cite{Du:2017qkv, Wu:2024gil},
\bqa
N^{eq}_{\Upsilon} (\tau) & = & V_{\rm FB} \gamma_b^2 n_\Upsilon(M_\Upsilon; T(\tau))  \nonumber \, , \\
& = & V_{\rm FB} \gamma_b^2 \frac{d_\Upsilon}{2 \pi^2} T M_\Upsilon^2 K_2\left(\frac{M_\Upsilon}{T} \right) \, ,
\label{eq:eqblmUpsilon}
\eqa
where the QGP fireball volume, $V_{\rm FB}$ (constructed from the aHydro simulations~\cite{Wu:2025lcj}), and bottom quark fugacity factor, $\gamma_b$, are implied to indirectly depend on the time $\tau$ through the relationship between $\tau$ and temperature and $K_2$ denotes the modified Bessel function of the second kind. The bottomonium state, with mass $M_\Upsilon$, has  degeneracy $d_\Upsilon$. 

In our calculations, the fireball volume is evaluated from the total entropy $S(\tau)$ and local entropy density $s(\tau)$ in three rapidity regions: central, $-1.93 < y < 1.93$; forward, $1.5 < y < 4.0$; and backward, $-5.0 < y < -2.5$. 
The total entropy in each region ${\mathcal Y} \in \{ \rm central, forward, backward \}$ is obtained on constant proper time slices according to 
\bqa
S_{\mathcal Y}(\tau) & = & 
\int_{\mathcal Y}\! d\eta_s \int dx\,dy\;\tau\, s_{\mathcal Y}\big(T(\tau,x,y,\eta_s)\big)\, \\ 
&\times &\Theta\big(T(\tau,x,y,\eta_s)-T_{\rm dec}\big) \, ,
\eqa
where $\Theta\big(T(\tau,x,y,\eta_s)-T_{\rm dec}\big)$  imposes the freeze-out temperature cut at $T_{\rm dec}$. 
The corresponding fireball volume is then
\begin{equation}
V_{\rm FB}^{\mathcal Y}(\tau) = \frac{S_{\mathcal Y}(\tau)}{s_{\mathcal Y}(\overline T_{\mathcal Y}(\tau))} \, ,
\end{equation}
with the average temperature in region $\mathcal Y$ defined as the arithmetic mean over the hydrodynamic fluid cells with temperatures greater than $T_{\rm dec} = 170$~MeV,
\begin{equation}
\overline  T_{\mathcal Y}(\tau) = \frac{1}{N_{\mathcal Y}(\tau)}
\sum_{\substack{(x,y,\eta_s)\in\mathcal Y \\ T>T_{\rm dec}}} T(\tau,x,y,\eta_s) \, ,
\end{equation}
where $N_{\mathcal Y}(\tau)$ counts the contributing trajectory samples at time $\tau$ in rapidity region $\mathcal{Y}$.

Consistent with the reaction rates, we employ the temperature dependent bottomonium masses, $M_\Upsilon (T)$, as obtained from the underlying $T$-matrix which is further related to the bottom quark mass, $m_b(T)$, and binding energy, $E_B(T)$, shown in Fig.~\ref{fig:mass}, as $M_\Upsilon (T) = 2 m_b(T) - E_B(T)$. The bottom quark fugacity is then calculated by imposing  $b\overline b$ pair conservation,
\beq
N_{b\overline b} (\tau) = N_{\rm op} (\tau) + N_{\rm hid} (\tau) \, ,
\label{eq:bbbar-number}
\eeq
where $N_{\rm op}$ and $N_{\rm hid}$ are the number of open ($b$ quarks and/or bottom hadrons) and hidden (bottomonia) bottom states, respectively, in the QGP. If the $b$ and $\overline b$ quarks are assumed to be uncorrelated, $N_{b\overline b}$ can be written in terms of the bottom quark thermal densities, 
\beq
n_{\rm b} (\tau) = \frac{d_b}{2\pi^2} T m_b^2 K_2\left(\frac{m_b}{T}\right) \, ,
\label{eq:n_Q}
\eeq
with $d_b$ = 6 for the $b$ quark (color-spin degeneracy) and hidden bottom states, $n_{\rm hid}$.  The fugacity factor, $\gamma_b$, can then be determined from 
\bqa
N_{b\overline b} (\tau) & = &\frac{1}{2}\gamma_b n_{\rm op}(\tau) V_{\rm FB}\frac{I_1(\gamma_b n_{\rm op} V_{\rm FB})}{I_0(\gamma_b n_{\rm op}  V_{FB})} \nonumber \\
& + & \gamma_b^2 (\tau) n_{\rm hid}(\tau) V_{\rm FB}(\tau) \, ,
\label{eq:bbbar-number1}
\eqa
where the quantity $I_1/I_0$ is the ratio of modified Bessel functions of the first kind.  

However, in practice, there are four important corrections to this calculation. First, in a strongly interacting QGP, strong correlations are expected to generate heavy-light resonance states which emerge from the in-medium $T$-matrices~\cite{Liu:2017qah} and have been shown to play an important role in the description of heavy quark susceptibilities~\cite{Liu:2021rjf}.  Thus, the thermal density of open bottom states in the system should be extended to include bottom hadrons, $n_{\rm op} = n_{b} + n_{H_b}$, where the second term in the sum is the density of the open bottom hadrons surviving in the sQGP. Since these are expected to ultimately melt at high temperatures,  
we construct a smooth transition for their disappearance based on the heavy-light $T$-matrices used in the rate calculations, parameterized as~\cite{Zhao:2010nk,Du:2017qkv}:
\beq
n_{\text{op}} = n_b + \frac{1}{2} \left\{ 1 - \tanh \left[ C_{H_b} \left( T - \overline T_{\rm diss} \right) \right] \right\} \sum n_{H_b},
\label{eq:n_open}
\eeq
where the open $B$ hadron density $n_{H_b}$, including both $B$ mesons and baryons is summed over the states listed in Table~\ref{tab:Hb}, along with their masses and degeneracies. We include the $S$-wave ground state open bottom mesons and baryons. We infer an average dissociation temperature of $\overline  T_{\rm diss}= 1.3 \times 170$~MeV, with a rather broad transition in temperature characterized by the coefficient $C_{H_b} =25$.  
\begin{table}[h!]
\centering
\begin{tabular}{|c|c|c|}
\hline
$B$ Meson & Mass (GeV) & Degeneracy \\
\hline
$B^{\pm}$ & 5.27931 & 2 \\
$B^0/\bar{B}^0$ & 5.27962 & 2 \\
$B^0_s/\bar{B}^0_s$ & 5.36682 & 2 \\
$B^{*\pm}$ & 5.32465 & 6 \\
$B^{*0}/\bar{B}^{*0}$ & 5.32465 & 6 \\
$B^{*0}_s/\bar{B}^{*0}_s$ & 5.41540 & 6 \\
\hline
   B Baryon & Mass (GeV) & Degeneracy \\
    \hline
    $\Lambda_b^0$            & 5.6196  & 2 \\
    $\Xi_b^0$                & 5.7919  & 2 \\
    $\Xi_b^-$                & 5.7970  & 2 \\
    $\Sigma_b^0$             & 5.8131  & 2 \\
    $\Sigma_b^+$             & 5.8106  & 2 \\
    $\Sigma_b^-$             & 5.8156  & 2 \\
      $\Sigma_b^{*0}$        & 5.8225  & 4 \\
    $\Sigma_b^{*+}$          & 5.8303  & 4 \\
    $\Sigma_b^{*-}$          & 5.8347  & 4 \\
    $\Lambda_b(5912)^0$      & 5.9122  & 2 \\
    $\Lambda_b(5920)^0$      & 5.9201  & 4 \\
    $\Xi_b^\prime(5935)^-$   & 5.9351  & 2 \\
    $\Xi_b^*(5945)^0$        & 5.9523  & 4 \\
    $\Xi_b^*(5955)^-$        & 5.9557  & 4 \\
    $\Omega_b^-$             & 6.0458  & 2 \\
    $\Lambda_b(6072)^0$      & 6.0723  & 2 \\
    \hline
\end{tabular}
\caption{Masses and degeneracies of the open bottom ground state mesons and baryons.}
\label{tab:Hb}
\end{table}

The second correction arises from the fact that in heavy-ion collisions $b\overline b$ production is essentially point-like so that the bottom quantum number needs to be conserved locally. Exact bottom flavor conservation is encoded in the ratio of Bessel functions in Eq.~\eqref{eq:bbbar-number1} and can be implemented by replacing the fireball volume, $V_{FB}$ with a finite correlation volume, $V_{\text{corr}} = \frac{4}{3} \pi (r_0 + \langle v_b \rangle \tau)^3$~\cite{Du:2017qkv}, over which the   $b\overline b$ pair can be located. Here $r_0 \simeq 0.8 - 1.2$~fm represents a typical strong interaction range while $\langle v_b \rangle \simeq (0.6 - 0.7)c$ is the recoil velocity of the $b$ quarks as they diffuse into the fireball volume.  
Thus Eq.~\eqref{eq:bbbar-number1} becomes 
\bqa
N_{b\overline b} (\tau) &= &\frac{1}{2}\gamma_b n_{\rm op}(\tau) V_{\rm FB} \frac{I_1(\gamma_b n_{\rm op} V_{\text{corr}})}{I_0(\gamma_b n_{\rm op}  V_{\text{corr}})} \nonumber \\
& + & \gamma_b^2 n_{\rm hid}(\tau) V_{\rm FB} \, .
\label{eq:bbbar-number2}
\eqa

The third correction accounts for the fact that the $b$ quarks are not thermally distributed in the  expanding fireball, tending to suppress the regeneration contribution. Following Ref.~\cite{Grandchamp:2002wp}, we thus rescale the   equilibrium limit computed from Eq.~\eqref{eq:eqblmUpsilon} with a relaxation time factor $\mathcal{R} = 1 - \exp{\left( \int_{\tau_0}^{\tau} d\tau'/\tau_b \right)}$ so that $N_{\rm eq} \xrightarrow{} \mathcal{R} \times N_{\rm eq}$. The $b$-quark thermalization time $\tau_b \sim 1/A(\langle p\rangle)$, where $A(p)$ is $b$-quark thermalization rate at its momentum $p$,
taken from the same $T$-matrix calculations constrained by the Wilson line correlators~\cite{Wu:2025hlf}, with $b$-quark average momentum $\langle p\rangle\approx 6\,$GeV.

The final correction comes from the fact that
$b$ quarks may exit active fireball volume before the temperature drops below $T_{\rm dec}$.
We estimate the fraction of $b\bar{b}$ pairs that remain inside the fireball at each time step,
\begin{equation}
f_{b\bar{b}}(t) = \frac{N_{b\bar{b}}^{\text{in}}(\tau)}{N_{b\bar{b}}} \ ,
\end{equation}
by sampling their initial positions from a Glauber-model distribution
and their transverse momentum $p_T$ from FONLL~\cite{Cacciari:1998it,Cacciari:2012ny,Cacciari:2015fta}, assuming back-to-back kinematics.
Multiplying this fraction to the total number of $b\bar{b}$ pairs yields the effective number of pairs present in the fireball at each time step,
\begin{equation}
N_{b\bar{b}}(\tau) \xrightarrow{}  f_{b\bar{b}}(\tau) \cdot N_{b\bar{b}}(\tau) \ .
\end{equation}
The bottom quarks that escape do not participate in bottomonium formation and therefore reduce the equilibrium numbers.


\subsection{$p_T$ spectra of regenerated bottomonia}
\label{ssec:coal}

We employ the instantaneous coalescence model~\cite{dover1991, Greco:2003mm} to calculate the $p_T$ spectra of bottomonia regenerated from $b$ and $\overline b$ quarks.  This model is appropriate for our purposes because of the long thermal relaxation times of the massive $b$ quarks.  In addition, the large bottomonia binding energies require that they be produced early in the QGP evolution~\cite{Du:2017qkv}. The number of bottomonia states formed via $b \overline b$ coalescence can be written as~\cite{Greco:2003mm}
\bqa
   N_\Upsilon & = & g_\Upsilon \int p_b \cdot d \Sigma_b p_{\overline b} \cdot d \Sigma_{\overline b}  \frac{d^3\textbf{p}_b}{(2\pi)^3E_b} \frac{d^3\textbf{p}_{\overline b}}{(2\pi)^3E_{\overline b}} \, \nonumber \\
   & \times & f_b(x_b,p_b) f_{\bar{b}}(x_{\overline b},p_{\overline b}) f_\Upsilon(x_b,x_{\overline b};p_b,p_{\overline b}) \, ,
\eqa
where $d \Sigma$ is an element of a space-like hypersurface with a temperature equal to the temperature at which the coalescence occurs.  The statistical factor $g_\Upsilon$ accounts for the internal quantum numbers required to form a color singlet bottomonium state from two spin-1/2 colored $b$ and $\overline b$ quarks. One has $g_\Upsilon = (2 l + 1)/12$ with $l = 0$ for $\Upsilon$($n$S) and $l=1$ for $\chi_b${($n$P) where the statistical factor of $1/12$ comes from the color and spin components~\cite{Greco:2003mm}. Note that here $x_b$ and $x_{\overline b}$ refer to the $b$ and $\overline b$ quarks location in position space.  The phase space distributions $f_b(x_b,p_b)$, $f_{\overline{b}}(x_{\overline b},p_{\overline b})$ of the $b$ and $\overline{b}$ are normalized to their corresponding multiplicities,
\beq
 N_{b,\overline{b}} =  \int p \cdot d \Sigma \frac{d^3\textbf{p}}{(2\pi)^3E} f_{b,\bar{b}}(x,p) \, .
\eeq
   
The functional form of the coalescence probability, $f_\Upsilon(x_b,x_{\overline b};p_b,p_{\overline b})$, depends on the overlap of the $b$ and $\overline{b}$ wavefunctions with the bottomonium wavefunction, as well as virtual parton emission required by four-momentum conservation. We assume that the position space and momentum space distributions of the $b$ quarks in the medium are uncorrelated and that, on average, the $b$ quark distributions are uniform in phase space.  Thus, the coalescence probability and $b$-quark phase space distributions can be reduced to momentum space only, \ie, $f(x,p) \propto f(p)$. A functional form that satisfies momentum conservation and wavefunction overlap is,
\beq
\label{eq:fUpsilon_def}
f_{\Upsilon}(x_b,x_{\overline b}; p_b,p_{\overline b})
\equiv
(2\pi)^3\,\delta^{(3)}\!\bigl(\mathbf{p_\Upsilon}-\mathbf{p}_b-\mathbf{p}_{\overline b}\bigr)\;
\omega (\mathbf{k})\,,
\eeq
where $\mathbf{k}$ is the relative momentum between the $b$ and $\overline b$ in the rest frame of the bottomonium state, $\mathbf{k}\equiv (\mathbf{p}_b-\mathbf{p}_{\overline b})/{\sqrt{2}}$. One potential form of the momentum space coalescence probability for formation of a bottomonium state with angular momentum $\ell$ is the Wigner function~~\cite{Sun:2017ooe,Greco:2003mm},
\beq
\omega(\mathbf{k})
= (2\pi)^{3} \frac{\bigl(4\pi\sigma_W^{2}\bigr)^{\tfrac{3}{2}}}{
(2\pi)^{3}}\frac{(2\sigma_W^{2}k^{2})^{\!\ell}}{(2\ell+1)!!}
\exp(-\sigma_W^2 \mathbf{k}^2),
\label{eq:wigner}
\eeq
where $\sigma_W$ is the width of the Wigner distribution.  
Equation~\eqref{eq:wigner} satisfies the normalization condition
\beq
\int d^3x\,d^3k\,\omega(\mathbf{k}) = (2\pi)^3.
\eeq
The widths of the S and P states can be expressed in terms of their radii,
\begin{align} 
\sigma_W^2(n{\rm S}) = \frac{4}{3} \langle r_{n{\rm S}}^2 \rangle \, \, , \, \, \, \sigma_W^2(n{\rm P}) = \frac{4}{5}\langle r_{n{\rm P}}^2 \rangle \, \, .
\label{eq:sigma}
\end{align}
The $b$-quark spectra
are obtained from fits to FONLL calculations for the initial spectra~\cite{Cacciari:1998it,Cacciari:2012ny,Cacciari:2015fta}, based on the assumption that $b$-quark rescattering effects in $p+A$ collisions can be neglected. With final-state bottomonium momentum of $\mathbf{p}_\Upsilon = \mathbf{p}_b + \mathbf{p}_{\overline b}$  the coalescence spectra are
\bqa
\frac{d^3N_{\Upsilon}^{\text{coal}}}{dyd^2p_T}
& = & C_{\text{reg}}\,g_{\Upsilon}
\frac{(4\pi\sigma_W^2)^{3/2}}{V_{\mathrm{FB}}}
\int d^3\mathbf{p}_b f_b(\mathbf{p}_b) f_{\overline b}(\mathbf{p}_\Upsilon - \mathbf{p}_b) 
\nonumber \\
& \times & \frac{(2\pi\sqrt{2})^3}{E_b E_{\overline b}}
\frac{(\sigma_W^2 (\mathbf{p}_\Upsilon - 2 \mathbf{p}_b)^2)^l}{(2l+1)!!} 
\nonumber \\
& \times & \exp(-\frac{1}{2}\sigma_W^2 (\mathbf{p}_\Upsilon - 2\mathbf{p}_b)^2) \, \, ,
\label{eq:coalspectra}
\eqa
where $C_{\text{reg}}$ is the normalization factor obtained from the solution of the rate equation, Eq.~(\ref{eq:rate-eq-reg-sol}), thus representing the regenerated bottomonia.

\subsection{$R^{\Upsilon}_{pA}$ including regeneration}
\label{ssec:Rpa-regen}
The nuclear modification factor due to bottomonia regeneration is
\beq
R^{\Upsilon(\rm reg)}_{pA} (y, p_T) =  \frac{d^2 N^{\Upsilon(\rm reg)}_{pA}/(d y \, d p_T)}{N_{\text{coll}} (d^2 N^{\Upsilon}_{pp})/(d y \, d p_T)}  \, .
\eeq
Feeddown from the regenerated bottomonia states must also be taken into account. Following Sec.~\ref{subsec:feeddown}, we have
\beq
R^{\Upsilon(\rm reg, incl)}_{pA} (y, p_T) = \frac{\left(F \cdot R^{\Upsilon(\rm reg)}_{pA}(p_T,y,\phi) \cdot \vec{\sigma}_{\text{direct}}\right)^{i}}{\vec{\sigma}_{\text{exp}}^{i}} \, .
\label{eq:feeddown_inreg}
\eeq

Finally, the total $R_{pA}^\Upsilon$ is defined as the sum of the primordial suppression, including both cold and hot matter effects, as shown in the previous section, and the regenerated contribution, 
\beq
R^\Upsilon_{pA} = R_{pA}^{\rm CNM} \times R^{\rm HNM (prim)}_{pA} + R^{\rm HNM (reg)}_{pA}  \, . \label{eq:rpa_tot}
\eeq

The uncertainty bands on the regeneration results were obtained by making the assumption that the $b$ and $\overline b$ quarks that become regenerated $\Upsilon$ states experience the same initial state nPDF effects as the bottomonium states.  While removing the restriction on the pair mass used to calculate the nPDF effects on $\Upsilon$ production may result in a somewhat weaker nPDF effect, see Ref.~\cite{Vogt:2019xmm}, this difference would have only a few percent effect on the regenerated yields, not enough to substantially affect our results. The uncertainty bands on the total $R_{pA}^\Upsilon$ in Eq.~\ref{eq:rpa_tot}) combine the nPDF and HNM uncertainties, added in quadrature.

We present and discuss our results including regeneration, combined with the CNM effects and primordial suppression presented earlier, as defined in Eq.~(\ref{eq:rpa_tot}) in the rest of this section. We focus on the TAMU-NP rates, as they produce the largest suppression and therefore, by detailed balance, they should also exhibit the greatest regeneration. The smaller TAMU-P rates would lead to a correspondingly smaller regeneration contribution. The  KSU+QTraj approach includes regeneration but only from the correlated $b \overline b$ pairs that would form bottomonium in $p+p$ collisions. 

\subsubsection{$\Upsilon$($n$S) regeneration}
\label{subsubsec:upsilon_reg}
\begin{figure*}[!hbt]
\begin{center}
\includegraphics[width=1.0\linewidth,height=4cm,keepaspectratio]{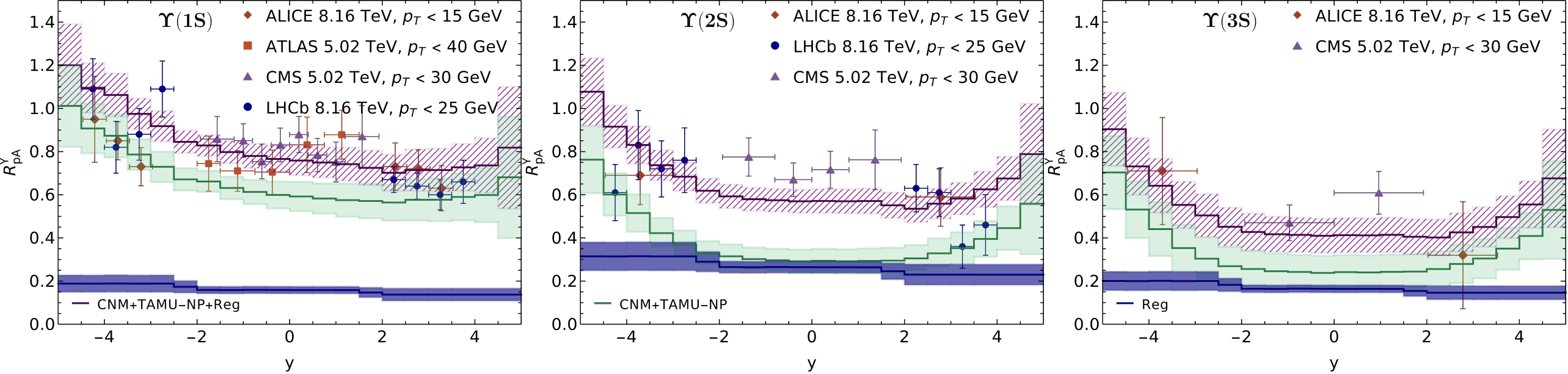}
\end{center}
\vspace{-4mm}
\caption{The total nuclear modification factor as a function of rapidity, including cold and hot nuclear matter suppression as well as regeneration, for $\Upsilon$(1S) (left), $\Upsilon$(2S) (middle) and $\Upsilon$(3S) (right). Results for CNM+TAMU-NP without (green band) and with (violet band) regeneration are compared. The regeneration-only $R_{\rm pA}$ is given by the dark blue bands. At $|y| < 2$ and $|y| \ge 2$, the results are shown for $\sqrt{s_{NN}} =  5.02$~TeV and $\sqrt{s_{NN}} = 8.16$~TeV, respectively.   The calculations are compared to data from ALICE~\cite{ALICE:2019qie}, ATLAS~\cite{ATLAS:2017prf}, CMS~\cite{CMS:2022wfi}, and LHCb~\cite{LHCb:2018psc}.
} 
\label{fig:rpA-y-Reg}
\end{figure*}

In Fig.~\ref{fig:rpA-y-Reg}, we present our results for $R_{pA}^{\Upsilon}$ as a function of rapidity $y$ for the $\Upsilon$(1S) (left), $\Upsilon$(2S) (middle), and $\Upsilon$(3S) (right) states.  At midrapidity, $|y|<2$, the results at $\sqrt{s_{NN}}=5.02$~TeV are compared to CMS~\cite{CMS:2022wfi} and ATLAS data~\cite{ATLAS:2017prf}, while at forward and backward rapidities, $|y|\gtrsim 2$, results at $\sqrt{s_{NN}}=8.16$~TeV are compared to ALICE~\cite{ALICE:2019qie}  and LHCb~\cite{LHCb:2018psc} data. As discussed previously, the CNM+TAMU-NP results alone generally give more suppression than seen in the data, particularly for the excited states.  However, the large suppression rates also allow for potentially significant regeneration. This effect is smallest on the $\Upsilon$(1S), $19$--$23\%$, improving agreement with the data. The regeneration contribution is stronger for the $\Upsilon$(2S), $25$--$32\%$, and $\Upsilon$(3S), $16$--$24\%$, bringing the results closer to the data.  Including regeneration  compensates for the stronger suppression, particularly for the excited states. 

The specific level of regeneration for each state depends on their equilibrium limit and the reaction rates.  These, in turn, depend on the thermal densities, meson masses and binding energies. Because the QGP is short lived in $p+$Pb collisions at the LHC, the interplay of all these results in the pattern $\Upsilon(1\rm S) \approx \Upsilon(3\rm S) < \Upsilon(2\rm S) $ for the regeneration contribution to $R_{pA}^{\rm HNM(reg)}$ alone.

\begin{figure*}[h!]
\begin{center}
\includegraphics[width=0.98\linewidth]{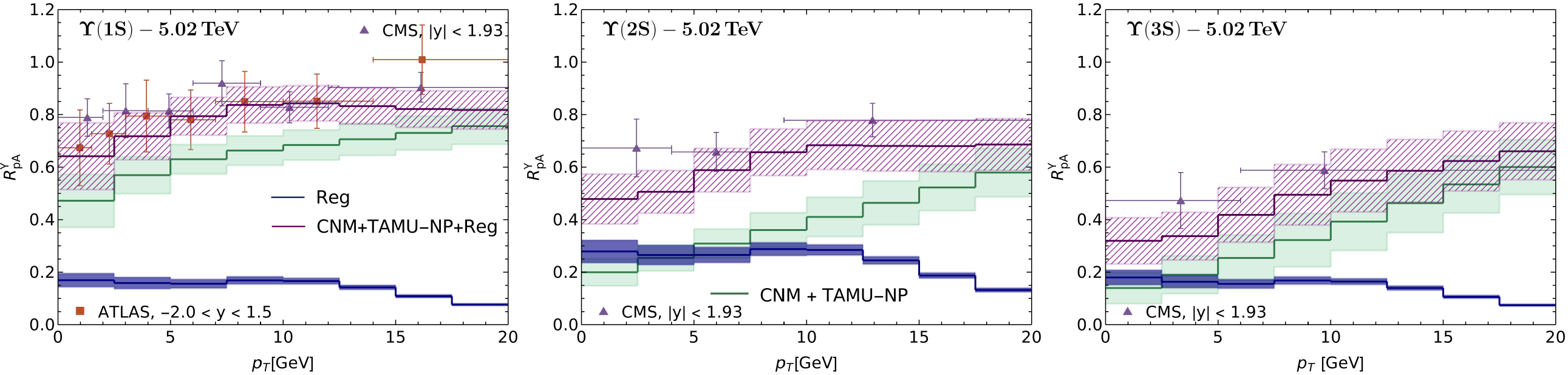}
\vspace{-5mm}
\end{center}
\caption{The nuclear modification factor $R_{pA}$ for $\Upsilon$(1S) (left), $\Upsilon$(2S) (middle), and $\Upsilon$(3S) (right) as a function of $p_T$ in $\sqrt{s_{NN}}=5.02$~TeV $p+$Pb collisions at $|y|<1.93$. The green band shows the CNM+TAMU-NP result without regeneration while the purple band includes regeneration (CNM+TAMU-NP+Reg). The calculation for regeneration alone is shown in the blue band.  The results are compared to CMS~\cite{CMS:2022wfi} and ATLAS~\cite{ATLAS:2017prf} data. The horizontal uncertainties indicate the $p_T$ bin widths.}
\label{fig:rpA-pt-nS-MidRap-5TeV-reg}
\end{figure*}

Figures~\ref{fig:rpA-pt-nS-MidRap-5TeV-reg} and \ref{fig:rpA-pt-farRap-8.16-withTAMUReg} illustrate our results as a function of $p_T$. 
Figure~\ref{fig:rpA-pt-nS-MidRap-5TeV-reg} gives the midrapidity results, $|y|<1.93$, for $\sqrt{s_{NN}}=5.02$~TeV compared to ATLAS~\cite{ATLAS:2017prf} and CMS~\cite{CMS:2022wfi} measurements. The effect of regeneration is largest at low $p_T$ and  decreases as $p_T$ increases.  The regeneration contribution  compensates for the substantial primordial suppression produced by large TAMU-NP reaction rates, filling the gap between the calculations with suppression alone and the data.

\begin{figure*}[h!]
\begin{center}
\includegraphics[width=0.98\linewidth]{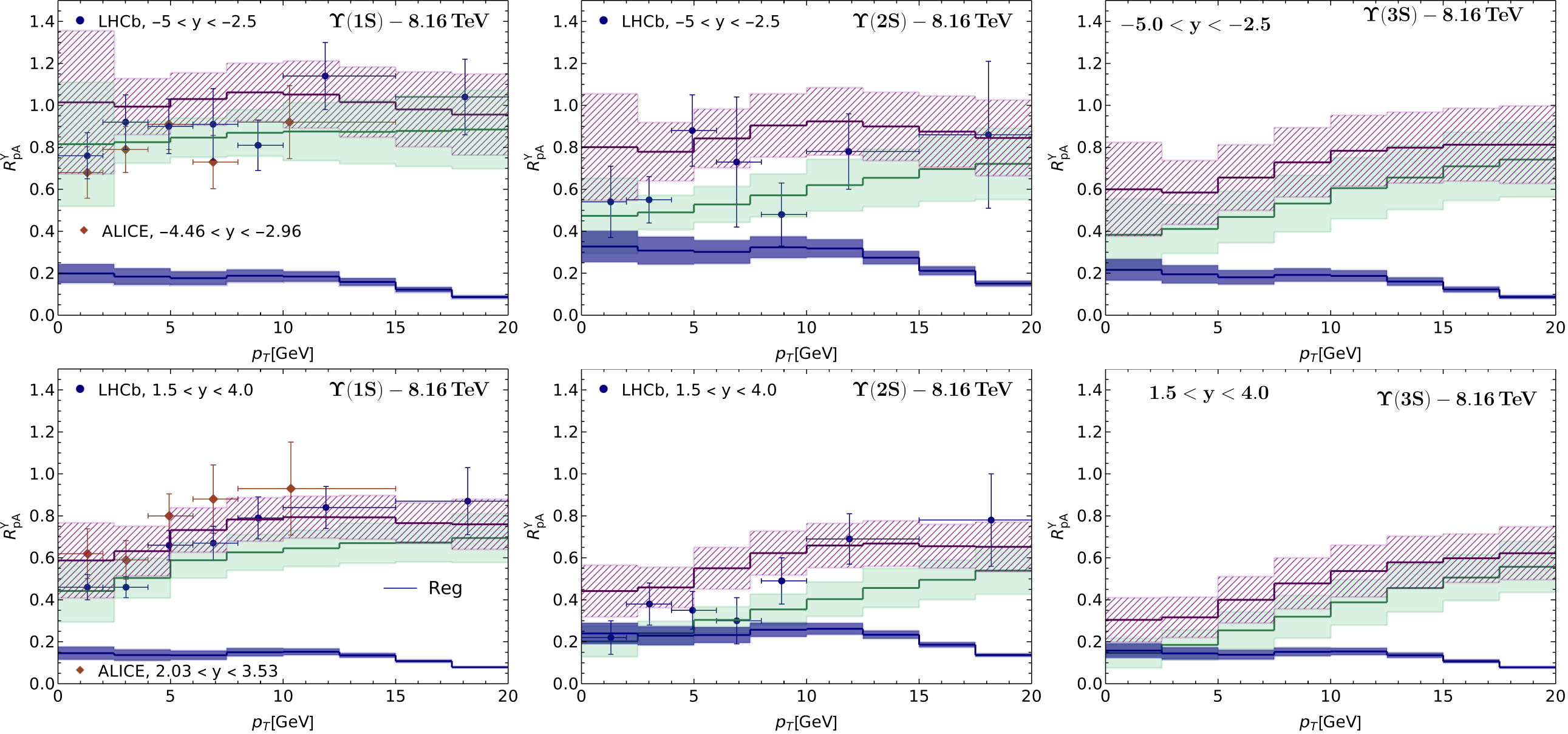}
\vspace{-5mm}
\end{center}
\caption{The nuclear modification factor $R_{pA}$ for $\Upsilon$(1S) (left), $\Upsilon$(2S) (middle), and $\Upsilon$(3S) (right)  as a function of $p_T$ in $\sqrt{s_{NN}}=8.16$~TeV $p+$Pb collisions at $-5.0 < y < -2.5$ (top row) and $1.4<y<4.0$ (bottom row). The green band shows the CNM+TAMU-NP result without regeneration while the purple band includes regeneration (CNM+TAMU-NP+Reg). The calculation for regeneration alone is shown in the blue band. The results are compared to data from ALICE~\cite{ALICE:2019qie} and LHCb~\cite{LHCb:2018psc}. The horizontal uncertainties indicate the $p_T$ bin widths.}
\label{fig:rpA-pt-farRap-8.16-withTAMUReg}
\end{figure*}

Figure~\ref{fig:rpA-pt-farRap-8.16-withTAMUReg} shows our results at $\sqrt{s_{NN}}=8.16$~TeV for backward rapidity, $-5.0<y<-2.5$ (top row) and forward rapidity,  $1.4<y<4.0$ (bottom row), compared to data from ALICE~\cite{ALICE:2019qie} and LHCb~\cite{LHCb:2018psc}. Even though the effect of regeneration is somewhat reduced away from midrapidity, it is still significant.  It is again strongest at low $p_T$, as expected.  However, an improvement in the agreement of the calculations with the data is not compelling.  The agreement is better for the $\Upsilon$(1S) at moderate $p_T$ at forward rapidity but including regeneration at backward rapidity seems to worsen the overall agreement as a function of $p_T$. The situation is even less clear for the $\Upsilon$(2S), due in part to the scatter in the data points. 

\subsubsection{$\chi_b$ regeneration}
\label{subsubsec:chistateswithReg}
Here, we present our predictions for the nuclear modification factor $R_{pA}$ of the $\chi_b$(1P) and $\chi_b$(2P) states in $\sqrt{s_{NN}} = 8.16$~TeV $p+{\rm Pb}$ collisions, including regeneration.

\begin{figure*}[ht!]
\begin{center}
\includegraphics[width=0.98\linewidth]{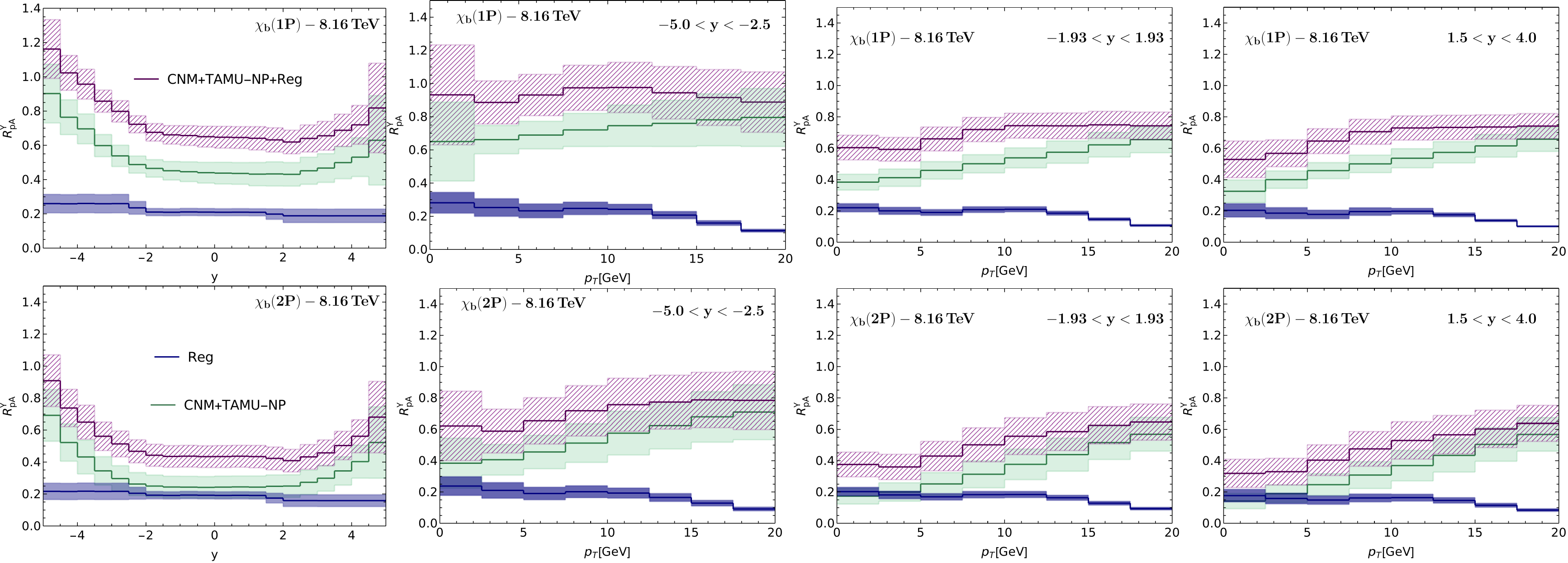}
\vspace{-5mm}
\end{center}
\caption{The nuclear modification factor $R_{pA}$ for $\chi_b$(1P) (top row) and $\chi_b$(2P) (bottom row) in $\sqrt{s_{NN}}=8.16$~TeV $p+{\rm Pb}$ collisions. Results are shown as functions of rapidity (left column) and $p_T$, in three rapidity regions: $-5.0<y<-2.5$ (center left); $|y|<1.93$ (center right); and $1.5<y<4.0$ (right). The CNM+TAMU-NP results are shown without (green band) and with (purple band) regeneration.  The regeneration contribution alone is shown by the blue bands.}  
\label{fig:rpA-xb-8TeV-withreg}
\end{figure*}

Figure~\ref{fig:rpA-xb-8TeV-withreg} presents our predictions for the nuclear modification factor of the $\chi_b$(nP) states as a function of rapidity $y$ (first column) and as a function of $p_T$ at backward rapidity (second column), midrapidity (third column), and forward rapidity (fourth column).
Significant regeneration is observed, similar to the $\Upsilon$($n$S) states.  The increase in $R_{pA}$ for the $\chi_b$(1P) is up to $\sim 20\%$ at midrapidity and up to $28\%$ at backward rapidity. 
In contrast, regeneration has a smaller effect on the $\chi_b$(2P), with only $10-22$\% increase in $R_{pA}$ due to regeneration.  

These results highlight the stronger sensitivity of $\chi_b$(2P) to the short-lived QGP medium. The $\chi_b$(1P) is more likely to be regenerated, despite its lower suppression rate.  Indeed, the regeneration-driven enhancement of $\chi_b$(1P) here aligns with the KSU+QTraj $\chi_b$ quantum-mechanical enhancement prediction, although it is not clear that the underlying mechanism is related.

\section{Conclusions and outlook}
\label{sec:conclusions}
We have performed a systematic study of bottomonium modification in minimum-bias $p+{\rm Pb}$ collisions relative to $p+p$ collisions at $\sqrt{s_{NN}} = 5.02$ and 8.16~TeV. Our analysis incorporates both cold and hot nuclear matter effects.  The cold matter effects include modifications of the nuclear parton distributions, coherent energy loss, and momentum broadening.  The hot matter effects arise due to final-state interactions in the quark-gluon plasma. We have employed two complementary approaches to model the QGP-induced suppression: the KSU-QTraj approach, based on open quantum systems, and the semiclassical TAMU approach. The dissociation rates in the TAMU framework include predictions based on the $T$-matrix $U$ potential with perturbative medium coupling (TAMU-P) and fully integrated nonperturbative $T$-matrix calculations with self-consistent off-shell spectral functions and constrained by recent Wilson line correlators (TAMU-NP). We have also studied the effect of bottomonium regeneration in $p+{\rm Pb}$ collisions for the first time.

Our results are presented in terms of the nuclear modification factor. They indicate that while cold matter effects establish a baseline suppression generally compatible with the $\Upsilon$(1S) data within uncertainties, hot matter effects are necessary to provide good descriptions of the $\Upsilon$(2S) and $\Upsilon$(3S) data. 

The modest hot matter effects on the $\Upsilon$(1S) are consistent among all three approaches. Differences emerge for the excited states.  The KSU-QTraj and the TAMU-P results are generally compatible with each other but stronger than that of cold nuclear matter alone. Significantly stronger suppression is observed for the TAMU-NP rates, particularly at forward and backward rapidity. 

Using the TAMU-NP rates, we found significant regeneration of the bottomonium states in the short-lived QGP, attributable primarily to the large dissociation rates. During the short time that the hot matter phase exists, the bottomonium states undergo rapid dissociation, increasing the regeneration probability.  Loosely bound states are found to be regenerated more frequently. The interplay between regeneration and primordial suppression can provide a reasonable description of the $\Upsilon$ data with the TAMU-NP rates.

We note that the same approach has recently been employed to study Pb+Pb collisions at the LHC~\cite{Wu:2025lcj}. Much like in the present work, without tuning any parameters in either the transport coefficients or in the background hydrodynamics, a fair description of the available LHC data was found. In a logical continuation of the results in the present paper, both suppression and regeneration processes are further reinforced.  Indeed, regeneration makes up nearly half of the $\Upsilon$(1S) production and dominates the excited states in semi-central Pb+Pb collisions. As such, a reasonably consistent picture has been obtained for bottomonium production from small to large collision systems within the framework of a strongly-coupled QGP, evolving hydrodynamically and featuring very large reaction rates for embedded heavy quarks.


We have also presented predictions for $\chi_b$ suppression in these collisions. Although data are not yet available for the \(\chi_b\) states, our predictions offer useful benchmarks for future measurements.

In the future, we plan to 
apply our approach to charmonium suppression in $p + {\rm Pb}$ and ${\rm d} + {\rm Au}$ collisions.  We will explore the contributions of QGP-induced dissociation and regeneration to charmonium suppression both as a function of the charmonium kinematics and collision centrality. 
These efforts will lead to a more comprehensive understanding of quarkonium behavior across a range of collision systems and energies.

\acknowledgments{
This work was supported by the U.S. Department of Energy, Office of Science, Office of Nuclear Physics through the Topical Collaboration in Nuclear Theory on {\it Heavy-Flavor Theory (HEFTY) for QCD Matter} under contract number DE-SC0023547. S.T. was supported by Kent State University through index 201452 and the Topical Collaboration in Nuclear Theory on Heavy-Flavor Theory (HEFTY) for QCD Matter under award no. DE-SC0023547. R.V.\ was also supported by the U.S. Department of Energy by LLNL under contract DE-AC52-07NA27344 and by the U.S. Department of Energy, Office of Science, Office of Nuclear Physics (Nuclear Theory) under contract number DE-SC-0004014. R.R.\ and B.W.\ were also supported by the U.S. National Science Foundation under grant nos.\ PHY-2209335 and PHY-2514775.  We would also like to thank Dr.\ Andrew Hanlon and Dr.\ Michael Strickland for valuable discussions on various aspects of this work.
}



\bibliographystyle{apsrev4-1}
\bibliography{main-inspire,main-notinspire}

\end{document}